\documentclass[twocolumn,preprint,astrosymb]{aastex631}
\usepackage{amsmath}

\shortauthors{McGrath et al.}

\begin{document}

\title{A Morphology Catalog of Galaxies in CEERS: Evolution in the Size and Color Gradients of Galaxies Since Cosmic Dawn}

\correspondingauthor{Elizabeth J. McGrath}
\email{emcgrath@colby.edu}

\author[0000-0001-8688-2443]{Elizabeth J.\ McGrath}
\affiliation{Department of Physics and Astronomy, Colby College, Waterville, ME 04901, USA}
\author[0000-0001-8519-1130]{Steven L. Finkelstein}
\affiliation{Department of Astronomy, The University of Texas at Austin, Austin, TX, USA}

\author[0000-0001-6813-875X]{Guillermo Barro}
\affiliation{Department of Physics, University of the Pacific, Stockton, CA 90340 USA}

\author[0000-0002-2499-9205]{Viraj Pandya}
\altaffiliation{Hubble Fellow}
\affiliation{Columbia Astrophysics Laboratory, Columbia University, 550 West 120th Street, New York, NY 10027, USA}
\author[0000-0001-7113-2738]{Henry C. Ferguson}
\affiliation{Space Telescope Science Institute, 3700 San Martin Dr., Baltimore, MD 21218, USA}

\author[0000-0001-9187-3605]{Jeyhan S. Kartaltepe}
\affiliation{Laboratory for Multiwavelength Astrophysics, School of Physics and Astronomy, Rochester Institute of Technology, 84 Lomb Memorial Drive, Rochester, NY 14623, USA}

\author[0000-0002-8360-3880]{Dale D. Kocevski}
\affiliation{Department of Physics and Astronomy, Colby College, Waterville, ME 04901, USA}

\author[0000-0001-5758-1000]{Ricardo O. Amor\'{i}n}
\affiliation{Departamento de Astronomía, Universidad de La Serena, Av. Juan Cisternas 1200 Norte, La Serena, Chile}
\affiliation{Instituto de Astrof\'{i}sica de Andaluc\'{i}a (CSIC), Apartado 3004, 18080 Granada, Spain}
\affiliation{ARAID Foundation. Centro de Estudios de F\'{\i}sica del Cosmos de Arag\'{o}n (CEFCA), Unidad Asociada al CSIC, Plaza San Juan 1, E--44001 Teruel, Spain}

\author[0000-0001-8534-7502]{Bren E. Backhaus}
\affiliation{Department of Physics and Astronomy, University of Kansas, Lawrence, KS 66045, USA}
\affiliation{Department of Physics, 196 Auditorium Road, Unit 3046, University of Connecticut, Storrs, CT 06269, USA}

\author[0000-0002-2861-9812]{Fernando Buitrago}
\affiliation{Departamento de F\'{i}sica Te\'{o}rica, At\'{o}mica y \'{O}ptica, Universidad de Valladolid, 47011 Valladolid, Spain}
\affiliation{Laboratory for Disruptive Interdisciplinary Science (LaDIS), Universidad de Valladolid, 47011 Valladolid, Spain}
\affiliation{Instituto de Astrof\'{i}sica e Ci\^{e}ncias do Espa\c{c}o, Universidade de Lisboa, OAL, Tapada da Ajuda, PT1349-018 Lisbon, Portugal}

\author[0000-0003-2536-1614]{Antonello Calabr\`o}
\affiliation{INAF Osservatorio Astronomico di Roma, Via Frascati 33, 00078 Monteporzio Catone, Rome, Italy}

\author[0000-0001-8551-071X]{Yingjie Cheng}
\affiliation{University of Massachusetts Amherst, 710 North Pleasant Street, Amherst, MA 01003-9305, USA}

\author[0000-0001-6820-0015]{Luca Costantin}
\affiliation{Centro de Astrobiolog\'{\i}a (CAB), CSIC-INTA, Ctra. de Ajalvir km 4, Torrej\'on de Ardoz, E-28850, Madrid, Spain}

\author[0000-0002-1803-794X]{Isa G. Cox}
\affiliation{Laboratory for Multiwavelength Astrophysics, School of Physics and Astronomy, Rochester Institute of Technology, 84 Lomb Memorial Drive, Rochester, NY
14623, USA}

\author[0000-0001-8047-8351]{Kelcey Davis}
\affiliation{Department of Physics, 196 Auditorium Road, Unit 3046, University of Connecticut, Storrs, CT 06269, USA}

\author[0000-0003-3248-5666]{Giovanni Gandolfi}
\affiliation{Dipartimento di Fisica e Astronomia "G.Galilei", Universit\'a di Padova, Via Marzolo 8, I-35131 Padova, Italy}
\affiliation{INAF--Osservatorio Astronomico di Padova, Vicolo dell'Osservatorio 5, I-35122, Padova, Italy}

\author[0000-0002-4162-6523]{Yuchen Guo}
\affiliation{Department of Astronomy, The University of Texas at Austin, Austin, TX, USA}

\author[0000-0001-6145-5090]{Nimish P. Hathi}
\affiliation{Space Telescope Science Institute, 3700 San Martin Dr., Baltimore, MD 21218, USA}

\author[0000-0002-3301-3321]{Michaela Hirschmann}
\affiliation{Institute of Physics, Laboratory of Galaxy Evolution, Ecole Polytechnique Fédérale de Lausanne (EPFL), Observatoire de Sauverny, 1290 Versoix, Switzerland}

\author[0000-0002-4884-6756]{Benne W. Holwerda}
\affil{Physics \& Astronomy Department, University of Louisville, 40292 KY, Louisville, USA}

\author[0000-0002-1416-8483]{Marc Huertas-Company}
\affil{Instituto de Astrof\'isica de Canarias, La Laguna, Tenerife, Spain}
\affil{Universidad de la Laguna, La Laguna, Tenerife, Spain}
\affil{Universit\'e Paris-Cit\'e, LERMA - Observatoire de Paris, PSL, Paris, France}

\author[0000-0002-6610-2048]{Anton M. Koekemoer}
\affiliation{Space Telescope Science Institute, 3700 San Martin Dr., Baltimore, MD 21218, USA}

\author[0000-0003-1581-7825]{Ray A. Lucas}
\affiliation{Space Telescope Science Institute, 3700 San Martin Dr., Baltimore, MD 21218, USA}

\author[0000-0001-5846-4404]{Bahram Mobasher}
\affiliation{Department of Physics and Astronomy, University of California, 900 University Ave, Riverside, CA 92521, USA}

\author[0000-0001-9879-7780]{Fabio Pacucci}
\affiliation{Center for Astrophysics | Harvard \& Smithsonian, 60 Garden Street, Cambridge, MA 02138, USA}
\affiliation{Black Hole Initiative, Harvard University, Cambridge, MA 02138, USA}

\author[0000-0001-7503-8482]{Casey Papovich}
\affiliation{Department of Physics and Astronomy, Texas A\&M University, College Station, TX, 77843-4242 USA}
\affiliation{George P.\ and Cynthia Woods Mitchell Institute for Fundamental Physics and Astronomy, Texas A\&M University, College Station, TX, 77843-4242 USA}

\author[0000-0003-4528-5639]{Pablo G. P\'erez-Gonz\'alez}
\affiliation{Centro de Astrobiolog\'{\i}a (CAB), CSIC-INTA, Ctra. de Ajalvir km 4, Torrej\'on de Ardoz, E-28850, Madrid, Spain}

\author[0000-0002-1410-0470]{Jonathan R. Trump}
\affiliation{Department of Physics, 196 Auditorium Road, Unit 3046, University of Connecticut, Storrs, CT 06269, USA}

\author[0000-0003-3466-035X]{{L. Y. Aaron} {Yung}}
\affiliation{Space Telescope Science Institute, 3700 San Martin Dr., Baltimore, MD 21218, USA}

%
%CEERS Architects not listed above
%
\author[0000-0002-7959-8783]{Pablo Arrabal Haro}
\altaffiliation{NASA Postdoctoral Fellow}
\affiliation{Astrophysics Science Division, Code 660, NASA Goddard Space Flight Center, 8800 Greenbelt Rd., Greenbelt, MD 20771, USA}

\author[0000-0002-9921-9218]{Micaela B. Bagley}
\affiliation{Astrophysics Science Division, Code 660, NASA Goddard Space Flight Center, 8800 Greenbelt Rd., Greenbelt, MD 20771, USA}

\author[0000-0001-5414-5131]{Mark Dickinson}
\affiliation{NSF's National Optical-Infrared Astronomy Research Laboratory, 950 N. Cherry Ave., Tucson, AZ 85719, USA}

\author[0000-0003-3820-2823]{Adriano Fontana}
\affiliation{INAF - Osservatorio Astronomico di Roma, via di Frascati 33, 00078 Monte Porzio Catone, Italy}

\author[0000-0002-5688-0663]{Andrea Grazian}
\affiliation{INAF--Osservatorio Astronomico di Padova, Vicolo dell'Osservatorio 5, I-35122, Padova, Italy}

\author[0000-0001-9440-8872]{Norman A. Grogin}
\affiliation{Space Telescope Science Institute, 3700 San Martin Dr., Baltimore, MD 21218, USA}

\author[0000-0001-8152-3943]{Lisa J. Kewley}
\affiliation{Center for Astrophysics | Harvard \& Smithsonian, 60 Garden Street, Cambridge, MA 02138, USA}

\author[0000-0002-5537-8110]{Allison Kirkpatrick}
\affiliation{Department of Physics and Astronomy, University of Kansas, Lawrence, KS 66045, USA}

\author[0000-0003-3130-5643]{Jennifer M. Lotz}
\affiliation{Space Telescope Science Institute, 3700 San Martin Dr., Baltimore, MD 21218, USA}

\author[0000-0001-8940-6768]{Laura Pentericci}
\affiliation{INAF - Osservatorio Astronomico di Roma, via di Frascati 33, 00078 Monte Porzio Catone, Italy}

\author[0000-0003-3382-5941]{Nor Pirzkal}
\affiliation{ESA/AURA Space Telescope Science Institute, 3700 San Martin Dr., Baltimore, MD 21218, USA}

\author[0000-0002-5269-6527]{Swara Ravindranath}
\affiliation{Astrophysics Science Division, NASA Goddard Space Flight Center, 8800 Greenbelt Road, Greenbelt, MD 20771, USA}
\affiliation{Center for Research and Exploration in Space Science and Technology II, Department of Physics, Catholic University of America, 620
Michigan Avenue N.E., Washington, DC 20064, USA}

\author[0000-0002-6748-6821]{Rachel S. Somerville}
\affiliation{Center for Computational Astrophysics, Flatiron Institute, 162 5th Avenue, New York, NY, 10010, USA}

\author[0000-0003-3903-6935]{Stephen M.~Wilkins} 
\affiliation{Astronomy Centre, University of Sussex, Falmer, Brighton BN1 9QH, UK}
\affiliation{Institute of Space Sciences and Astronomy, University of Malta, Msida MSD 2080, Malta}

\author[0000-0001-8835-7722]{Guang Yang}
\affiliation{Kapteyn Astronomical Institute, University of Groningen, P.O. Box 800, 9700 AV Groningen, The Netherlands}
\affiliation{SRON Netherlands Institute for Space Research, Postbus 800, 9700 AV Groningen, The Netherlands}

%
% Other authors
%
\author[0000-0001-7755-4755]{Lise-Marie Seillé}
\affiliation{Aix Marseille Univ, CNRS, CNES, LAM, Marseille, France}

\author[0000-0002-9373-3865]{Xin Wang}
\affiliation{School of Astronomy and Space Science, University of Chinese Academy of Sciences (UCAS), Beijing 100049, China}
\affiliation{National Astronomical Observatories, Chinese Academy of Sciences, Beijing 100101, China}
\affiliation{Institute for Frontiers in Astronomy and Astrophysics, Beijing Normal University, Beijing 102206, China}

%\suppressAffiliations

\begin{abstract}

We present measurements of morphological parameters from fitting 53,885 galaxies detected to a magnitude limit of F356W$< 28.5$ in the CEERS NIRCam imaging with {\sc galfit} in six broadband filters: F115W, F150W, F200W, F277W, F356W, and F444W. We provide a public catalog of S\'ersic index, effective semi-major axis, axis ratio, integrated magnitude, and position angle for these galaxies in each of the filters. Uncertainties in the measured parameters are estimated from simulated galaxies that 
have similar noise and background properties as the observed galaxies. We compare our measurements with those in the CANDELS/EGS field measured with HST/WFC3 and find that the sizes agree to within 0.09 dex and the S\'ersic indices agree to within 0.13 dex. We further present the evolution in the size-mass relation, and find that the evolution to $z\sim9$ is consistent with previous results derived at lower redshift. Finally, we look at the color gradients of galaxies at $1<z<5$ and find that for late-type galaxies ($n<2.5$), there is a strong dependence on mass, but no apparent evolution with redshift, indicating that the stellar populations and dust attenuation in more massive galaxies vary substantially with radius and contribute to significant morphological $k-$corrections.  For early type galaxies ($n>2.5$), the color gradients are nearly flat with no dependence on mass, indicating that the stellar populations are more uniform throughout.
The structural measurements presented are accurate to $20\%$ or better for most galaxies with F356W $<27.0$~mag and will enable further studies of galaxy morphology to $z\sim10$.

\end{abstract}

\section{Introduction} \label{sec:intro}
Statistical samples of galaxy morphology across cosmic time are useful in constraining formation and evolutionary schemes.
In the past several decades, key observations have revealed that 
the observed bimodality in galaxy color correlates with structure (\citealt{hubble36,holmberg58,devauc61,roberts94,strateva01}), 
that galaxies were smaller in the past (\citealt{ferguson04,papovich05,trujillo06,buitrago08,vanderWel08,williams10,mosleh12,cassata13,ono13,vanderWel14,allen17,mowla19,nedkova21,suess22,baggen23,ormerod24,ward24}), that the size-mass relation is dependent on morphological type (\citealt{shen03, vanderWel14, lange15, casura22}), and that the Hubble sequence was already in place at early times (\citealt{lee13,jacobs23,ferreira23,kartaltepe23,huertas-company23}).

A key era where galaxy properties are rapidly changing is ``cosmic noon'' at $z\sim2$ \citep{madau14}. Linking galaxy morphology at earlier times to the present day requires higher resolution data than ground-based, seeing-limited observations can provide. Furthermore, in order to probe the underlying mass distribution of galaxies, observations need to be carried out at rest-frame optical wavelengths, or longer, and ideally in the same rest-wavelength across all redshifts to minimize morphological k-corrections. Beyond $z\sim1.5$, this requires observations at infrared (IR) wavelengths. To date, the largest gains in the understanding of galaxy structure and its evolution during cosmic noon have come from the \emph{Hubble Space Telescope} (HST) and large IR surveys made with the WFC3 instrument (e.g., CANDELS: \citealt{koekemoer11, grogin11}, 3D-HST: \citealt{vanDokkum11, brammer12, skelton14}, 
and COSMOS-DASH: \citealt{mowla19, cutler22}).
However, even these surveys cannot probe beyond $z\gtrsim3$ at rest-optical wavelengths, where the $4000$\AA-break falls beyond the edge of the F160W filter. Additionally, due to the observed size evolution, galaxies at $z>2$ are poorly resolved in much of the HST imaging of the previous decade, especially more compact sources, including lower-mass and quiescent galaxies.  JWST improves on both of these issues, allowing rest-frame optical imaging of galaxies to $z\sim10$ with a spatial resolution of a few hundred parsecs at the highest redshifts.

The Cosmic Evolution Early Release Science (CEERS) survey \citep{finkelstein22,bagley23,ceers25}
is one of many recent wide-field imaging surveys with JWST/NIRCam. CEERS is an early release science program that covers $\sim$90 arcmin$^{2}$ of the Extended Groth Strip (EGS; \citealt{davis07}) with imaging and spectroscopy using coordinated, overlapping parallel observations by most of the JWST instrument suite. 
As an early release science program, data from the field became available immediately after observation for use by the scientific community. Several papers involving morphological studies of galaxies have already been published using CEERS data (e.g., \citealt{suess22,ferreira23,kartaltepe23,nelson23, huertas-company23, costantin23, sun24, vega-ferrero24, ormerod24, pandya24, ward24, vanderWel24, allen25}). 
JWST has allowed the study of rest-frame optical and rest-frame near-IR morphologies at $z>2$ in a host of other fields, as well (e.g., \citealt{gillman24,shuntov25,costantin25,yang25}).
Many of these previous works focused on mass- or redshift-limited samples, often defined by prior HST imaging and generally limited to a few thousand sources.
In this paper, we create a morphology catalog for 53,885 galaxies detected over the full CEERS area to a limiting magnitude of F356W$<28.5$ and provide a public release of the dataset to enable further studies of galaxy structural evolution in the era of JWST.
Throughout this paper we use AB magnitudes and assume a flat cosmology with $H_0 = 70$~km~s$^{-1}$~Mpc$^{-1}$, $\Omega_M = 0.3$, and $\Omega_{\Lambda}=0.7$.

\section{Data Description} \label{sec:data}
CEERS is based around a mosaic of 10 NIRCam pointings, with six NIRSpec and eight MIRI pointings observed in parallel.  
In each NIRCam pointing, data were obtained in the short-wavelength (SW) channel F115W, F150W, and F200W filters, and long-wavelength (LW) channel F277W, F356W, F410M, and F444W filters.  The total exposure time for pixels observed in all three dithers was typically 2835 s per filter.
We use the publicly available reductions of the CEERS data\footnote{CEERS data is available at MAST: \dataset[doi: 10.17909/Z7P0-8481]{\doi{10.17909/Z7P0-8481}}.} (v0.5) as described in \citet{bagley23}, which include custom processing steps to perform snowball correction, wisp subtraction, $1/f$ noise removal, and background subtraction. The final mosaics have been drizzled to a pixel scale of 0\farcs03/pixel.

As noted in \citet{bagley23}, the background subtraction algorithm masks source flux in successive tiers to account for both extended and compact sources, and grows the mask generously around all galaxies before fitting the unmasked pixels with a two-dimensional model. This method, like any background subtraction algorithm, must impose upper limits on the size of sources to mask, resulting in some diffuse light in the wings of the largest galaxies ($\gtrsim 2\farcs4$ in radius) being erroneously subtracted.
The smoothing scale of the background fitting procedure (roughly 50 pixels, or $1\farcs5$) produces a very uniform background that is consistent on large scales with the rms expected from a perfectly flat image affected only by the counting statistics of the incoming photons. We note, however, that any real structures that were not masked would be subtracted by this method. Given the uniformity of the background after this fitting procedure, we apply no further background fitting when determining best-fit galaxy models. For large galaxies, where the wings have been suppressed by this background subtraction, their resulting morphology could be affected, however for these sources there are likely other complexities that are not well encapsulated by a single S\'ersic model (e.g., spiral arms, bulges, bars, clumps, etc.). 

\subsection{Photometry}\label{sec:photometry}
Photometry was computed on point-spread-function (PSF)-matched images using SExtractor \citep{bertin96} version~2.25.0 in two-image mode, with an inverse-variance weighted combination of the PSF-matched F277W and F356W images as the detection image, as described in \citet{finkelstein23}.  Photometry was measured in all seven of the NIRCam bands observed by CEERS. For the purpose of deriving photometric redshifts and masses (see \S\ref{photometric-redshifts} below), we also include photometry from the six HST bands (F606W, F814W, F105W, F125W, F140W, F160W), as described in \citet{finkelstein23}. We note that the photometry catalog and detection characteristics were optimized for faint, compact sources at high redshift. 
This ``hot" mode detection has the drawback of splitting (or segmenting, in SExtractor parlance) bright and large galaxies into two or more components. We note therefore that morphological parameters of bright galaxies derived using these detection parameters should be checked against other pre-existing measurements, such as those from the CANDELS/EGS field with HST/WFC3 \citep{vanderWel12} (see also, \S \ref{sec:candels_comparison}). Even given these limitations, \citet{pandya24} note that our catalog finds the majority of sources detected with an independent run of SExtractor++ (SE++; \citealt{bertin20,kummel22}), and that the main differences are for large and bright galaxies, where SE++ finds slightly more sources, due to the fact that these galaxies are segmented into multiple smaller components in our catalog.

\subsection{Photometric Redshifts and Masses}
\label{photometric-redshifts}
We measure photometric redshifts and rest-frame colors for all sources in our 13-band photometric catalog using EAZY \citep{brammer08}. We run EAZY with two different setups. In the first setup, we use our fiducial Kron, aperture-corrected photometry with a maximum redshift of 20, and a customized template list as described in \citet{finkelstein23}, which includes bluer templates from \citet{larson23} that are better-suited to the highest-redshift sources. We assume a flat prior in luminosity, include a systematic error of 5\% of the observed flux values, and fit to our measured total flux and flux error values.
In the second setup, we use the more recent python version of the code EAZYpy with the default template set “tweak fsps QSF 12 v3” which consists of a set of 12 templates derived from the stellar population synthesis code FSPS \citep{conroy10}. 

In addition, we also estimate stellar population properties by fitting the optical and NIR spectral energy distributions (SEDs) using FAST \citep{kriek09}, assuming \citet{bc03} stellar population synthesis models, following a \citet{chabrier03} initial mass function (IMF), a delayed exponential star formation history (SFH), and the \citet{calzetti00} dust law with attenuation $0<A_{V} <4$ mag.

\section{{\sc galfit} Measurements} \label{sec:galfit}
{\sc galfit} \citep{peng10} v3.0.5 was run individually on the CEERS NIRCam mosaics in each of the six broadband filters (F115W, F150W, F200W, F277W, F356W, and F444W).
We fit single-component S\'ersic profiles to sources with F356W $< 28.5$ mag, and used values from the SExtractor catalog as starting parameters for position, magnitude ($m$), effective radius ($r_e$), position angle ($P.A.$), and axis ratio ($q$). We set the initial guess for S\'ersic index to $n=2.5$. Empirical PSFs, generated by stacking isolated stars across all CEERS fields \citep{finkelstein23}, were used as input to {\sc galfit}.
For single-S\'ersic profiles, {\sc galfit}'s $\chi^2$-minimization is fairly robust when starting from initial guesses for the parameters measured directly from the data, as we do here. More complex methods for evaluating best-fit parameters, including allowing a wider range of input parameters and performing a full Markov Chain Monte Carlo analysis (e.g., \citealt{lange16}) are not feasible for our sample size.

In order to generate thumbnail images to be fit, we used 10 times the default SExtractor Kron ellipse (i.e., A\_IMAGE $\times$ KRON\_RADIUS, Kron parameter $k=2.5$), but placed an upper limit on the fitting region of 10\farcs0.  This upper limit encloses 2 times the traditional Kron ellipse (or roughly 15 times the half-light radius) 
for 98\% of all sources. 
We used matching thumbnail cutouts of the ERR array as input noise maps (i.e., ``sigma images'') for each source in {\sc galfit}. The ERR array includes both Poisson noise from the source (VAR\_POISSON) and a rescaled VAR\_RNOISE to match the sky variance, as described in \citet{bagley23}. 

The same detection image was used for all filters (see \citealt{finkelstein23} for details). In fitting primary sources with F356W $< 28.5$ mag, we used the following criteria to determine how to treat other sources that fell within the image thumbnail. All sources within the fitting region brighter than 27th magnitude in the filter of interest and no more than three magnitudes fainter than the primary source were fit simultaneously. This means that faint neighboring sources are sometimes treated differently in each filter, but we find that the choice of masking or fitting these faint sources has little impact on the primary galaxy fit.
The choice of 27th magnitude for neighboring sources, rather than fitting all neighbors up to 28.5 mag, was made not only to save computational time, but also because {\sc galfit} often fails to converge on a solution when there are too many free parameters, such as when there are too many galaxies fit simultaneously.
Galaxies whose centroids fell outside the fitting region, those whose magnitudes were $>3$ magnitudes fainter than the primary source, or galaxies with magnitudes greater than 27th magnitude were masked during fitting, using the segmentation map to define the masked region. 
During fitting, we held the background value fixed at zero and we placed the following constraints on parameters to keep the fit within reasonable bounds: x and y centroid within 3 pixels of the input value, Sersic index $0.2\leq n \leq 8.0$, effective radius $0.009 \leq r_e \leq 12.0$~arcsec, axis ratio $0.01 \leq q \leq 1$, and magnitude within 3 mag of the input SExtractor value.

\begin{figure}
    \centering
    \includegraphics[scale=0.3]{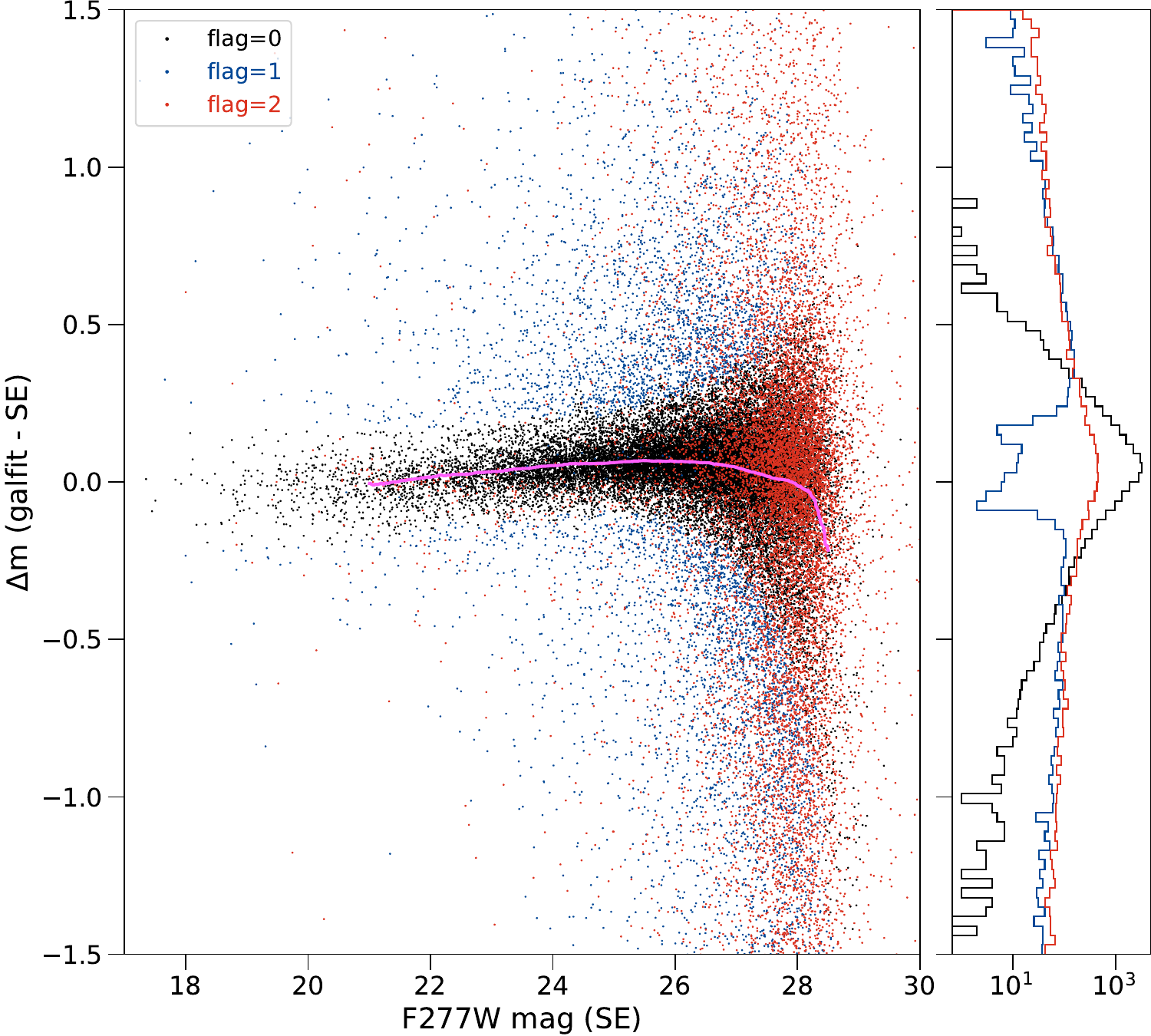}
    \caption{Left: Difference between F277W {\sc galfit} model magnitudes and SExtractor magnitudes for galaxies where the {\sc galfit} solution converged successfully (89\% of all sources). The magenta line shows a running median of the offset between the two magnitudes.  Galaxies with good fits (flag=0) are shown in black. Galaxies whose {\sc galfit} magnitude is more than $3\sigma$ away from the median offset (flag=1) are shown in blue.  Galaxies whose fits reached a constraint limit in one or more parameters (flag=2) are shown in red. Right: Histogram of sources as a function of difference between F277W {\sc galfit} magnitude and SExtractor magnitude. Colors are the same as in the left-hand panel.}
    \label{fig:galfit-se}
\end{figure}

\begin{figure*}
    \centering
    \includegraphics[scale=0.95]{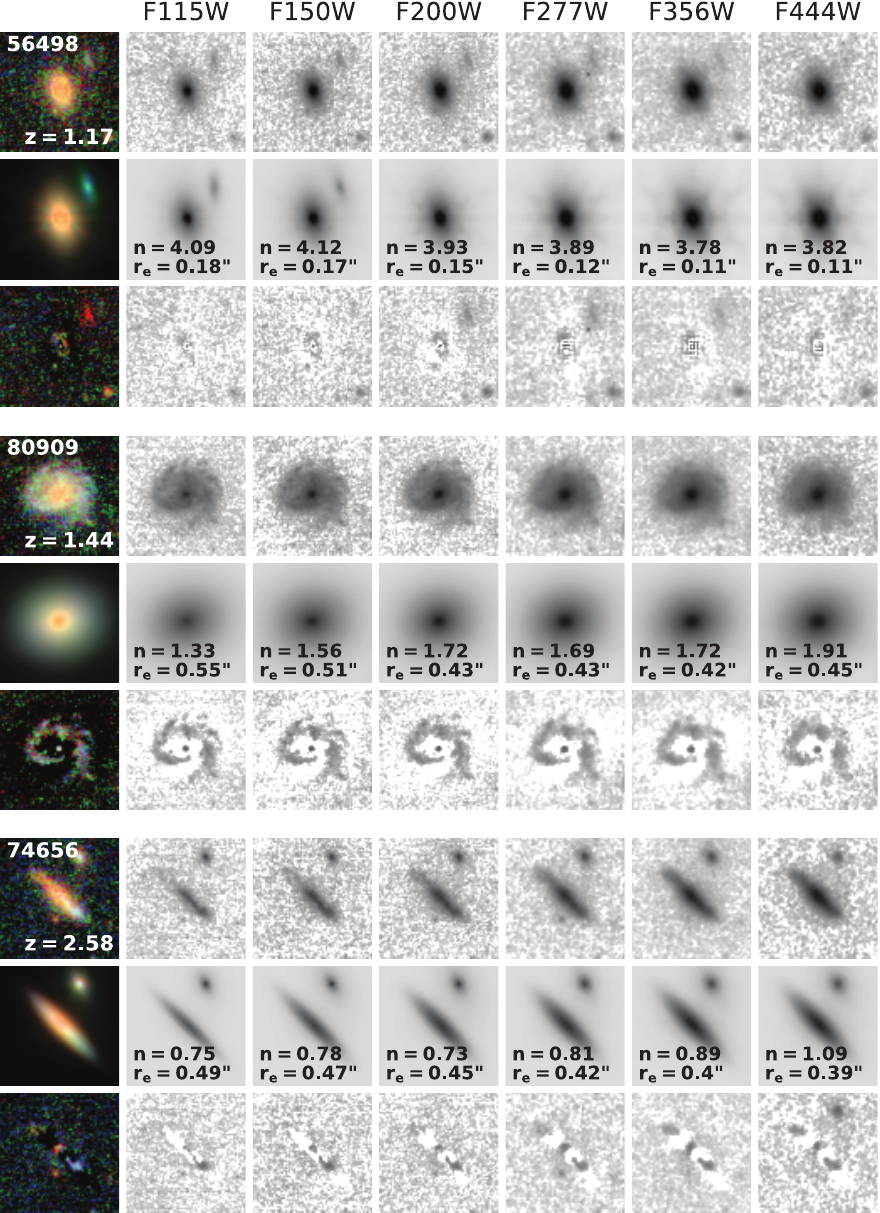}
    \vspace{-0.0in}
    \caption{An illustration of the {\sc galfit} fitting in six broadband filters for three galaxies of different morphological types. Images are 3\farcs0 on a side. Results for each galaxy are shown in three rows; top row: a color composite followed by images of the galaxy in each individual filter, middle row: model color composite followed by the galfit model in each filter, bottom row: residual color composite followed by the residuals obtained by subtracting the best-fit model in each filter. 
    The galaxy ID in the CEERS catalog and its photometric redshift are listed in the color composite image panel. The best fitting S\'ersic index $n$ and effective semi-major axis $r_e$ for each filter are listed in the middle row for each model. CEERS-56498 (top) is an elliptical galaxy, well-fit by an $n\sim4$ profile in all bands. CEERS-80909 (middle) is a face-on disk whose spiral arms are noticeable in the residuals. CEERS-74656 is an edge-on disk with a dust-lane that is noticeable in the transition from blue to red wavelengths, and results in uneven residuals.}
    \label{fig:galfit-residuals}
\end{figure*}

After running {\sc galfit}, sources were flagged according to the quality of the fit. Flags are assigned for each filter independently. Sources with reliable fits 
were defined as those whose fits converged and whose model magnitude fell within 3 times the dispersion, $\sigma$, in the running median offset between {\sc galfit} and SExtractor magnitudes (magenta line, Fig.\ref{fig:galfit-se}). These were assigned 
a flag value of 0 (roughly 45\% of all sources). Sources whose best-fit magnitudes 
were greater than $3\sigma$ away from the running median offset between the {\sc galfit} and SExtractor magnitude
were given a flag value of 1 (roughly 10\% of all sources).
Sources with a flag value of 2 are sources where one or more parameters reached a constraint limit during fitting (roughly 35\%). Sources with a flag value of 3 are sources that failed to fit (roughly 10\%; known simply as ``bombs'' in {\sc galfit}), generally because of too many neighboring galaxies that were being fit simultaneously.
Finally, sources with a flag value of 4 (roughly 0.3\% of sources) are 
artifacts in the photometry catalog that were not fit by {\sc galfit}.
A selection of fits for galaxies of different morphological types 
are shown in Fig.~\ref{fig:galfit-residuals}.

\section{Simulated Data and Measurement Errors}
\label{sec:errors}
\subsection{Methodology}
\label{sec:error_method}
The parameter errors produced by {\sc galfit} are generally underestimated (e.g., \citealt{haussler07}). In order to determine reliable errors for morphological parameters, we simulated 10,000 galaxies with {\sc galfit} with a range of parameters following a similar distribution in $m$, $n$, $r_e$, $q$, and $P.A.$ to observed galaxies in CEERS. Specifically, we used a log-normal distribution in $m$, $n$, and $r_e$, with $20\leq m \leq 28.5$, $0.5 \leq n \leq 8$, and $0\farcs05 \leq r_e \leq 1\farcs0$, along with a smooth distribution in $q$ and $P.A.$, with $0.1 \leq q \leq 1.0$, and $-90^{\circ} \leq P.A. \leq 90^{\circ}$. 
These simulated galaxies were generated in the F200W filter and convolved with the same empirical PSF described in \S \ref{sec:galfit} and then placed into blank regions of the CEERS mosaic so that they have the same background noise properties as real galaxies, but no additional Poisson noise was added to the sources. When assigning input magnitudes to sources, we ensured that the box size was large enough that when the model galaxy profile was integrated, we recover the input magnitude. While there are some observed galaxies outside of the parameter range used for generating simulated galaxies, the maximum brightness and radii were chosen so that galaxies would fit within blank sky regions with minimal source overlap. 
We ran our source detection on these images (see \S \ref{sec:data}), which recovered 8069 unique galaxies from the initial 10,000. We used this photometry catalog of 8069 sources to get initial parameters and segmentation maps for use with {\sc galfit}. Then we ran {\sc galfit} on these simulated galaxies in the same manner as for the observations (see \S \ref{sec:galfit}), including or masking neighbors (both simulated and real) in the fit, where appropriate. Of the 8069 input galaxies, 461 failed to fit (flag=3) and 1334 reached one or more constraint limits during fitting (flag=2), resulting in a final catalog of 6274 simulated sources with fits that converged and did not reach a constraint limit (i.e., flag$<2$). 
For the analysis that follows, we only consider the 6274 simulated galaxies with good fits.

Using this dataset, we derive empirical errors for the observed galaxies following a similar methodology as \citet{vanderWel12}. For each observed galaxy in the CEERS imaging, we locate the 100 most similar galaxies recovered in the simulated dataset. 
As in previous work \citep{vanderWel12}, measurement errors are assumed to be dependent only on $m$, $n$, and $r_e$, and thus, we use these parameters to determine a galaxy's similarity to another in this 3D parameter space. We do this comparison between target galaxy parameters and simulated galaxy parameters in logarithmic intervals so that differences between galaxies in each parameter correspond to fractional differences. Furthermore, we divide each parameter by its standard deviation in order to produce dimensionless, normalized quantities.  Thus, for each target galaxy in CEERS we determine a 3D ``distance" in logarithmic parameter space to each simulated galaxy:

\begin{multline}
\label{eqn:dist}
d_{i,j} = \sqrt{ \left(\frac{(m_i - m_j)}{\sigma(m)} \right)^2 + \left(\frac{(\log n_i - \log n_j)}{\sigma(\log n)} \right)^2 } \\  + \left(\frac{(\log r_{e,i} - \log r_{e,j})}{\sigma(\log r_e)} \right)^2 
\end{multline}

where $d_{i,j}$ is the normalized parameter distance from observed galaxy $i$ to simulated galaxy $j$, and $m_j$, $\log n_j$, and $\log r_{e,j}$ are the recovered parameters for the simulated galaxies. 
Then, since we know truth a priori for the simulated galaxies, we take the 100 simulated galaxies with the closest distance to the observed galaxy and compute the lower 16th and upper 84th percentile range in the difference (in logarithmic units) between recovered value and truth for each parameter of interest, which includes $m$, $n$, $r_e$, $q$, and $P.A.$. We assume errors are Gaussian distributed and assign 1$\sigma$ empirical errors to each observed galaxy as one-half this full spread for each parameter.

Simulated images were created for the F200W filter, but we perform a similar error analysis for the other filters, using the SNR to properly scale the errors from one filter to another as described in \citet{vanderWel12}. Having derived empirical errors for all galaxies in F200W, we then use this dataset as a starting point for the other filters. For a given target galaxy in a given filter, we use the 25 most similar galaxies in F200W as determined by equation \ref{eqn:dist}, multiply each error in F200W by the SNR in F200W, take the average value, and then divide by the SNR of the target galaxy in the filter of interest.  In multiplying by the ratio of the SNR in F200W to the SNR in the filter of interest, we generate errors with the proper amplitude for data at different depths (due to varying exposure times and wavelength-dependent throughput).
While the PSF also differs between filters, and resolution effects could have additional systematic effects that are not encapsulated here, we do not expect that resolution differences will affect the derived random uncertainties significantly.  We do find that the choice of PSF impacts the recovered parameters to some extent (see \S \ref{sec:psf-effects}), and that the errors are correlated with the morphological parameters themselves (see \S \ref{sec:errors-correlated}), but that both of these factors are small, suggesting that the combined effect would likely be overshadowed by more significant contributors, such as SNR. We tested this assumption with a smaller subset of simulated galaxies (1,519) where we produced images in the F356W filter, as well, and found that the spread in (recovered - truth) for a given range in $m$, $n$, and $r_e$ was similar to that found in the F200W images.

\subsection{Correlation Between Errors}
\label{sec:errors-correlated}
\begin{figure*}
    \centering
    \includegraphics[scale=0.70]{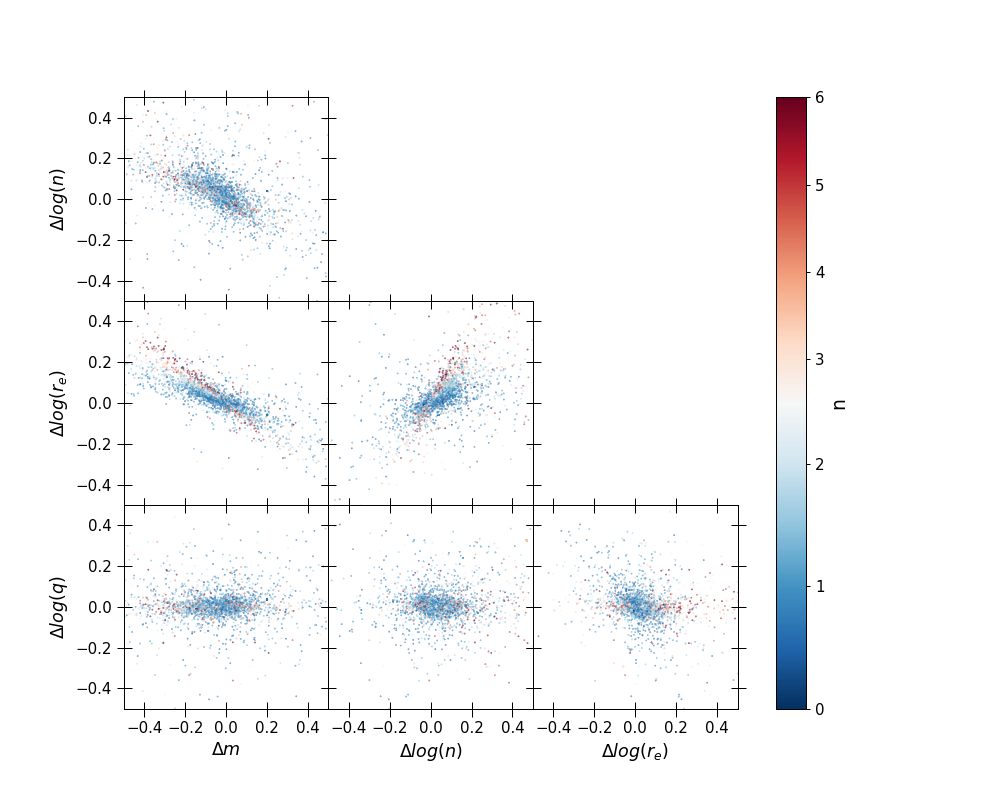}
    \caption{Correlation between errors for simulated galaxies, presented as the difference between (recovered - input) values for each parameter of interest: $m$, $n$, $r_e$, and $q$. Objects whose recovered magnitudes are brighter than truth tend to have larger recovered sizes and S\'ersic indices (upper left and middle left panels). Likewise, objects with larger recovered $n$ tend to have larger recovered $r_e$ values (middle panel). There is no obvious correlation between errors in axis ratio with the other parameters (bottom row).}
    \label{fig:errors-correlated}
\end{figure*}

It is well known that uncertainties are correlated between various morphological parameters (e.g., \citealt{haussler07, guo09, vanderWel12}), as well as the parameters themselves being correlated with each other (e.g., $n$ and $r_e$ as in \citealt{graham96}). In Figure \ref{fig:errors-correlated}, we show the difference in recovered - true values for a set of parameters ($m$, $n$, $r_e$, and $q$), which represents the measurement error for each simulated galaxy. 
We find that, in particular, errors in recovered $n$, $r_e$, and $m$ are correlated with each other, while there is no obvious correlation for any of the uncertainties with errors in $q$ (or $P.A.$, not shown). Objects whose recovered magnitudes are brighter than truth tend to have larger recovered sizes and S\'ersic indices. This intuitive behavior arises because higher $n$ results in more compact profiles, and thus the $r_e$ of the model needs to increase in order to fit extended flux at larger radii. Likewise, integrating over a larger $n$ (and $r_e$) causes the flux for the source to be overestimated, as well. 

An additional complicating factor is that the errors are also correlated with the morphological parameters themselves. We show one such correlation with S\'ersic index, $n$, in Figure~\ref{fig:errors-correlated}, through use of the color-coding by $n$. It is clear that sources with higher $n$ also have larger errors in recovered $r_e$, which is reflected in the middle panels as well as the lower right panel of Figure ~\ref{fig:errors-correlated}, where clear offsets can be distinguished between the errors for low-$n$ and high-$n$ sources.

\subsection{PSF Effects}
\label{sec:psf-effects}
Derived morphological parameters are dependent on accurate model point spread functions (PSFs). To evaluate the effects of uncertainty on the measured parameters due to the assumed PSF, we created a model PSF with WebbPSF, which differs from the empirical PSF that was used to create the simulated galaxy profiles (described in section \ref{sec:error_method}). The size of the WebbPSF was 10\farcs0 compared with 3\farcs0 for the empirical PSF, and was resampled to a pixel scale that was four times oversampled with respect to the CEERS mosaic images. No dithering or drizzling was applied to match the CEERS observations. 
This PSF was then used to fit the simulated galaxies with {\sc galfit} and recovered parameters were compared with those obtained using the empirical PSF.

We find that the median ($\pm$ the median absolute deviation) for the recovered $n$ and $r_e$ 
are smaller using the WebbPSF by 
$0.03\pm0.02$~dex and $0.01\pm0.01$~dex, respectively, while the recovered $q$ is larger by $0.01\pm0.02$~dex, and the recovered magnitude is brighter by $0.04\pm0.02$~mag.
Small magnitude offsets are likely due to the different treatment of flux in the wings of the empirical versus WebbPSF, while differences in $n$ and $r_e$ are correlated, as expected (\S \ref{sec:errors-correlated}). 

These values reflect the median difference in parameters using the two different PSFs along with their median absolute deviation, however there are additional dependencies on $n$ and $r_e$, as well. 
Galaxies with $n<2.5$ (i.e., the majority of all galaxies) have recovered $n$ values that are 
$0.03\pm0.02$~dex smaller with WebbPSF, $r_e$ values that are similar ($0.00\pm0.01$~dex), $q$ values that are $0.01\pm0.01$~dex larger, and magnitudes that are $0.04\pm0.01$~mag brighter.
Galaxies with $n>2.5$ have recovered $n$ and $r_e$ values that are 
$0.10\pm0.04$~dex and $0.04\pm0.03$~dex smaller with WebbPSF, respectively, and $q$ that is $0.02\pm0.02$~dex larger, but magnitudes that are $0.01\pm0.04$ mag fainter.
Similarly, for the smallest galaxies with $r_e<0\farcs18$, $n$ is
$0.07\pm0.04$~dex smaller, while $r_e$ and $q$ are $0.01\pm0.01$~dex and $0.04\pm0.03$~dex larger with WebbPSF, respectively, and magnitudes are $0.05\pm0.01$ mag brighter.
For larger galaxies with $r_e>0\farcs18$, $n$ and $r_e$ are 
$0.03\pm0.02$~dex and $0.01\pm0.01$~dex smaller, respectively, while $q$ is $0.01\pm0.01$~dex larger, and magnitudes are $0.03\pm0.02$~mag brighter.
Given that the differences in recovered $n$ and $r_e$ values are most pronounced for $n>2.5$ and $r_e<0\farcs18$ galaxies, we emphasize that an accurate PSF is most important for small, high-S\'ersic-index galaxies.

\citet{sun24} also studied the effects of using WebbPSF versus empirical PSFs derived directly from the data using NIRCam imaging of a globular cluster (M92). They found that the empirical PSFs performed significantly better in fitting stars in the field (resulting in better $\chi^2_{\nu}$) and in recovering the true stellar flux.  Thus, while differences in the recovered parameters for high-$n$, low-$r_e$ galaxies is more dependent on choice of PSF, we expect that empirical PSFs will give the best results overall.

Finally, \citet{pandya24b} compared our measurements of galaxy ellipticity and orientation to those of individual stars throughout CEERS for a gravitational lensing analysis. They found that the CEERS PSF varies spatially with an average ellipticity of $\sim2\%$ in the bluer NIRCam filters and $\lesssim1\%$ in the redder NIRCam filters. These spatial variations are roughly consistent with our global empirical PSF from stacking the stars and are generally too small to bias the shape measurements except for the smallest or roundest galaxies.

\begin{figure*}
    \centering
    \includegraphics[scale=0.40]{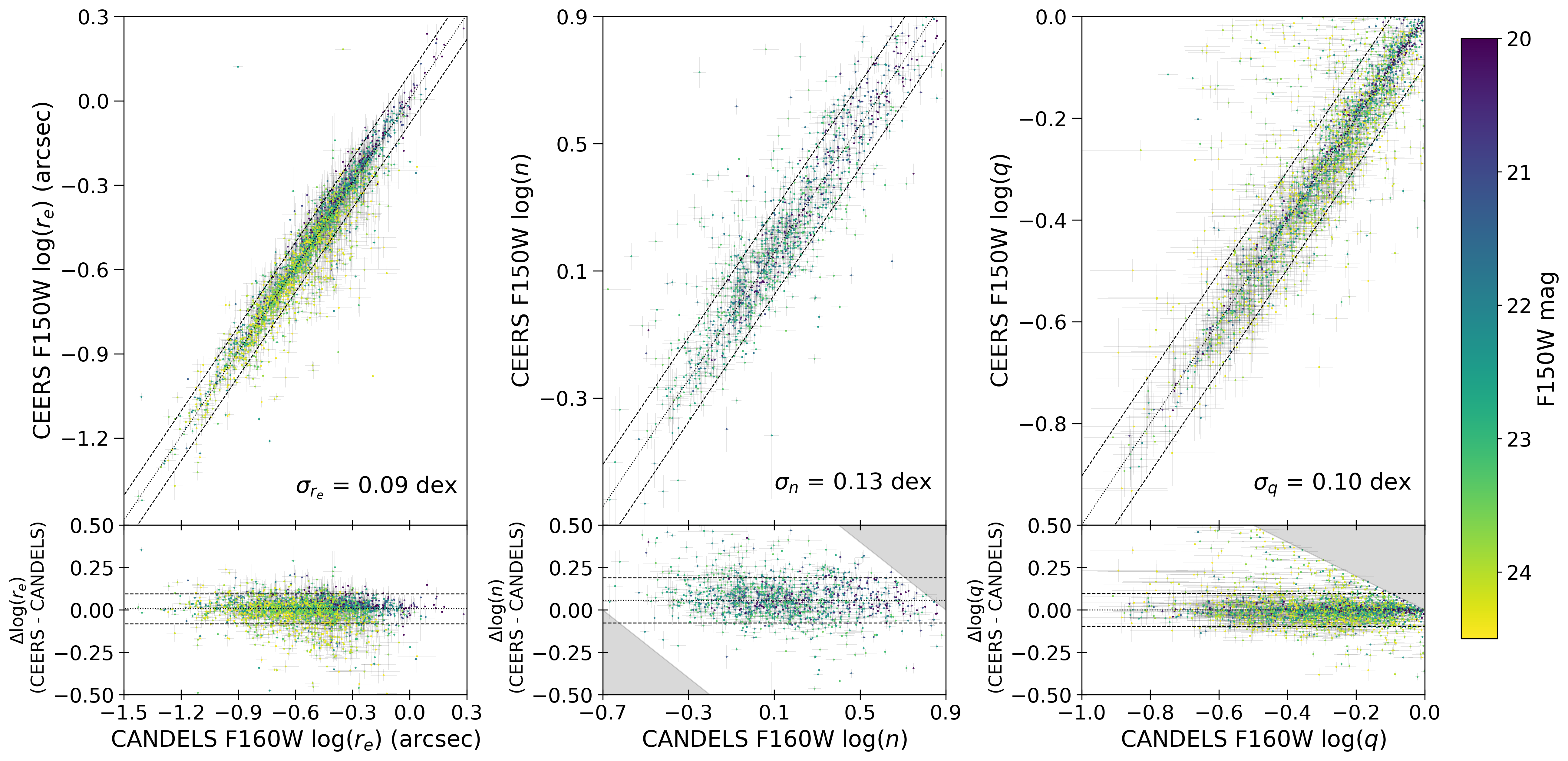}
    \caption{From left to right we show a comparison between measured effective radii, $r_e$; S\'ersic indices, $n$; and axis ratios, $q$ for galaxies detected in both the CEERS and CANDELS surveys, restricted to galaxies with $m_{\rm{F150W}} < 24.5$ ($r_e$ and $q$) or $m_{\rm{F150W}} < 23.5$ ($n$), where the CANDELS values are quoted to have random uncertainties less than $\sim20\%$ \citep{vanderWel12}. CEERS values are from JWST/NIRCam F150W images, while CANDELS datapoints are from HST/WFC3 F160W images \citep{vanderWel12}. The top panels show the 1:1 comparison between surveys while the bottom panels show the offset (in log space) of the CEERS values from CANDELS. The 1:1 correspondence is shown as a dotted line in each panel, along with the $1\sigma$ spread shown as the upper and lower dashed lines whose magnitude is quoted in the lower right corner of the top panels. 
    For $n$, there is a noticeable offset in the best-fit value between CEERS and CANDELS, with CEERS $n$ being larger by 0.055 dex (i.e., $14\%$ larger in linear units), and we have offset the 1:1 line and corresponding spread by this amount accordingly. There is a slightly less noticeable offset between CEERS and CANDELS $r_e$ values, with CEERS $r_e$ being 0.01 dex larger over this magnitude range. We have similarly offset the 1:1 line and corresponding spread by this value in the left-hand panels. Shaded regions in the lower middle and right panels indicate either unphysical regions of parameter space, or regions disallowed by a constraint set on a parameter in one or both datasets.}
    \label{fig:candels_comparison}
\end{figure*}

\section{Results} \label{sec:results}

In Table \ref{tab:galfit_catalog}, we provide our final measurements, including $m$, $n$, $r_e$, $q$, and $P.A.$, along with empirical errors. Note that the effective radius is an effective semi-major axis, rather than a circularized effective radius. For brevity, we show only the first five lines using results from F277W, and note that the full dataset along with similar measurements for the other filters are available in the online version of the journal.
For non-logarithmic quantities (e.g., $n$, $r_e$, $q$, $P.A.$), we quote the errors in linear units as $\delta p = \ln(10)p\delta \log p$, where $p$ is the parameter of interest and $\delta \log p$ is the 1$\sigma$ error determined as described in \S \ref{sec:error_method}. For magnitudes, we note that we give best-fit model magnitudes from {\sc galfit}, rather than the SExtractor magnitude from the input photometry. Thus, while we used F356W$<28.5$ as the magnitude limit for fitting, some sources are recovered with model magnitudes fainter than this limit. 

Note that the unique IDs listed in Table \ref{tab:galfit_catalog} correspond to our own detection and photometry (as described in section \ref{sec:photometry}). For ease of use with ancillary data products, however, we perform a nearest neighbor match (within 0\farcs3) with the official CEERS photometry catalog (Cox et al. 2025, submitted) and also provide this CEERS ID, where available, for each source.  Since the detection algorithms differ, there is not a one-to-one correspondence for all sources. The Cox et al.~catalog follows a ``hot+cold'' algorithm that combines smaller sub-components of larger galaxies into a single detection (see Cox et al. 2025, in preparation, for details).  We expect that for typical sources with $r_e < 1\farcs0$, the source detection and segmentation should be similar, and thus the physical parameters presented in the official CEERS catalog can be used alongside the morphology parameters presented herein.

In Table \ref{tab:percentile_errors} we list the magnitudes that result in size ($r_e$) and shape ($n$) errors better than 20\% for 50-, 75-, 90-, and 95-percent of the data. Limiting magnitudes for similar fractional errors in magnitude ($m$), axis ratio ($q$), and position angle ($P.A.$) are not shown, but they are generally much fainter since these parameters are better determined than either $r_e$ or $n$. Based on this, we suggest the following magnitude limits for highest-fidelity of the morphology catalog values presented herein: $m_{\rm{F115W}}\leq26.5$, $m_{\rm{F150W}}\leq26.25$, $m_{\rm{F200W}}\leq26.5$, $m_{\rm{F277W}}\leq26.75$, $m_{\rm{F356W}}\leq27.0$, and $m_{\rm{F444W}}\leq26.4$.
We recommend the use of the published morphological values for sources with flag values $<2$ falling within these magnitude ranges for the user's filter of interest for greatest reliability. 
In \S$5.1-5.3$, we compare our results with previous work, and preview some science results using these data.

\begin{deluxetable*}{cccccccccc}
    \tablewidth{0pt}
    \label{tab:galfit_catalog}
    \centering
    \tablecaption{{\sc galfit} Results in F277W}
    \tablehead{
         \colhead{ID} & \colhead{CEERS ID$^{\dag}$} & \colhead{RA}  &  \colhead{Dec}  & 
         \colhead{$m \pm \delta m$}  &  \colhead{$r_e \pm \delta r_e$}  &  \colhead{$n \pm \delta n$}  &  
         \colhead{$q \pm \delta q$}  &  \colhead{P.A. $\pm \delta$P.A.}  &  \colhead{FLAG} \\
        \colhead{}  & \colhead{}  & \colhead{J2000}  &  \colhead{J2000}  &  \colhead{AB mag}  &  \colhead{$^{\prime\prime}$}  & \colhead{}   & \colhead{}   &  \colhead{deg}  & \colhead{}
        }
    \startdata
        1  & 17343 &  214.983215  &  52.952023  &   $19.74\pm 0.04$  &  $0.44 \pm 0.02$  &  $1.13 \pm 0.06$  &  $0.49 \pm 0.01$  &  $8.06 \pm 2.08$  &  0  \\
        2  & 17346 &  214.984985  &  52.951199  &   $19.61\pm 0.06$  &  $0.33 \pm 0.02$  &  $2.72 \pm 0.19$  &  $0.52 \pm 0.01$  & $-71.38 \pm 1.15$  &  0  \\
        3  & 18121 &  215.034012  &  52.986191  &   $19.60\pm 0.05$  &  $0.57 \pm 0.02$  &  $1.39 \pm 0.07$  &  $0.17 \pm 0.01$  &  $44.05 \pm 2.14$  &  0  \\
        4  & \nodata &  214.947769  &  52.980446  &   $23.68\pm 1.66$  &  $12.0 \pm 5.29$  &  $6.50 \pm 9.66$  &  $0.24 \pm 0.39$  &  $31.70 \pm 50.61$  &  2  \\
        5  & 50378 &  214.950714  &  52.982414  &   $23.13\pm 0.04$  &  $0.17 \pm 0.01$  &  $3.44 \pm 0.25$  &  $0.43 \pm 0.01$  &  $81.05 \pm 1.62$  &  0  \\
        .  & . &  .  &  .  &  .  &  .  &  .  &  .  & .  &  . \\
        .  & . &  .  &  .  &  .  &  .  &  .  &  .  & .  &  . \\
        .  & . &  .  &  .  &  .  &  .  &  .  &  .  & .  &  . \\[0.1in]
    \enddata
    \tablenotetext{}{Results for the entire sample of 53,885 galaxies in all six filters (F115W, F150W, F200W, F277W, F356W, and F444W) are available to download as a single FITS table at this site: https://github.com/ejmcgrath/ceers-galfit .}
    \tablenotetext{\dag}{From Cox et al. 2025, submitted}
\end{deluxetable*}

\begin{deluxetable*}{c|cccc|cccc}
    \tablewidth{0pt}
    \label{tab:percentile_errors}
    \centering
    \tablecaption{Magnitude limits corresponding to percentile of measurements with fractional errors $<20\%$}
    \tablehead{ \colhead{} & \multicolumn{4}{c}{$\delta r_e < 20\%$} & \multicolumn{4}{c}{$\delta n < 20\%$} \\ 
    \colhead{Filter} & \colhead{50\%} & \colhead{75\%}  &  \colhead{90\%}  & \colhead{95\%}  & \colhead{50\%}  &  \colhead{75\%}  &  \colhead{90\%}  &  \colhead{95\%} }
    \startdata
       F115W &  27.25  & 26.42   &  25.65   & 25.23   &    26.82  & 26.12 & 25.43 & 25.00 \\
       F150W &  26.97  & 26.16   &  25.37   & 24.92   &    26.53  & 25.79 & 25.19 & 24.81 \\
       F200W &  27.22  & 26.42   &  25.84   & 25.41   &    26.81  & 26.13 & 25.58 & 25.32 \\
       F277W &  27.52  & 26.67   &  26.10   & 25.71   &    27.11  & 26.38 & 25.92 & 25.56 \\
       F356W &  27.71  & 26.86   &  26.31   & 25.92   &    27.36  & 26.63 & 26.16 & 25.83 \\
       F444W &  27.10  & 26.29   &  25.68   & 25.34   &    26.75  & 26.04 & 25.49 & 25.19 \\
    \enddata
    \tablenotetext{}{Magnitude limits are calculated using flag$<2$ sources.}
\end{deluxetable*}

\subsection{Comparison to Previous Work}
\label{sec:candels_comparison}
The CEERS NIRCam footprint overlaps CANDELS \citep{grogin11,koekemoer11} coverage with HST/WFC3. Using the public {\sc galfit} catalog from CANDELS \citep{vanderWel12}, we compare fits for galaxies detected in both datasets in Fig.~\ref{fig:candels_comparison}. Galaxies in the two datasets were matched based on the nearest neighbor within a 0\farcs3 radius. We find a match for 18,270 sources. Since detection algorithms differ between CANDELS and this work, some galaxies may be segmented differently, which could affect the resulting morphological measurements. If, for example, one CANDELS galaxy was split into two CEERS sources, the resulting morphology may be dominated by different substructures in the two catalogs. With this limitation in mind (which generally only affects the brightest sources), we compare structural measurements in F160W (CANDELS) and F150W (CEERS) for galaxies with magnitudes where the CANDELS measurements are expected to have random uncertainties of $20$\% or better for $r_e$ and $q$ (F160W$<24.5$~mag), or similar accuracy for $n$ (F160W$<23.5$~mag), as discussed in \citet{vanderWel12}. We limit our comparison to galaxies with flag=0 in F160W and F150W filters in each catalog (2503 and 1281 sources, respectively, for the two different magnitude limits, and $z<4.5$).

As can be seen in Fig.~\ref{fig:candels_comparison}, there is excellent agreement between the CEERS F150W and CANDELS F160W morphologies, derived from datasets with differing sensitivities, resolutions, and pixel scales. 
The $1\sigma$ spread in the difference between $r_e$ values in the two catalogs for sources in this magnitude range
is only 0.09 dex.  For $n$, the $1\sigma$ difference is 0.13 dex, and for $q$ it is 0.10 dex. 

While the agreement between CEERS and CANDELS morphological parameters is generally good, we note that there is a systematic offset between the best-fit $n$ for the same galaxies in both datasets. In the middle panel of Fig.~\ref{fig:candels_comparison}, we find that the 1:1 correspondence line is offset by 0.055 dex, with CEERS data recovering a higher $n$ than CANDELS. Offsets such as this could be due to differences in SNR (see below), or background subtraction (e.g., \citealt{haussler07}). Likewise, the average $r_e$ in CEERS is 0.01 dex larger than CANDELS (Fig.~\ref{fig:candels_comparison}, left panel). 

\begin{figure}
    \centering
    \includegraphics[scale=0.40, trim={0 0cm 0 0cm},clip]{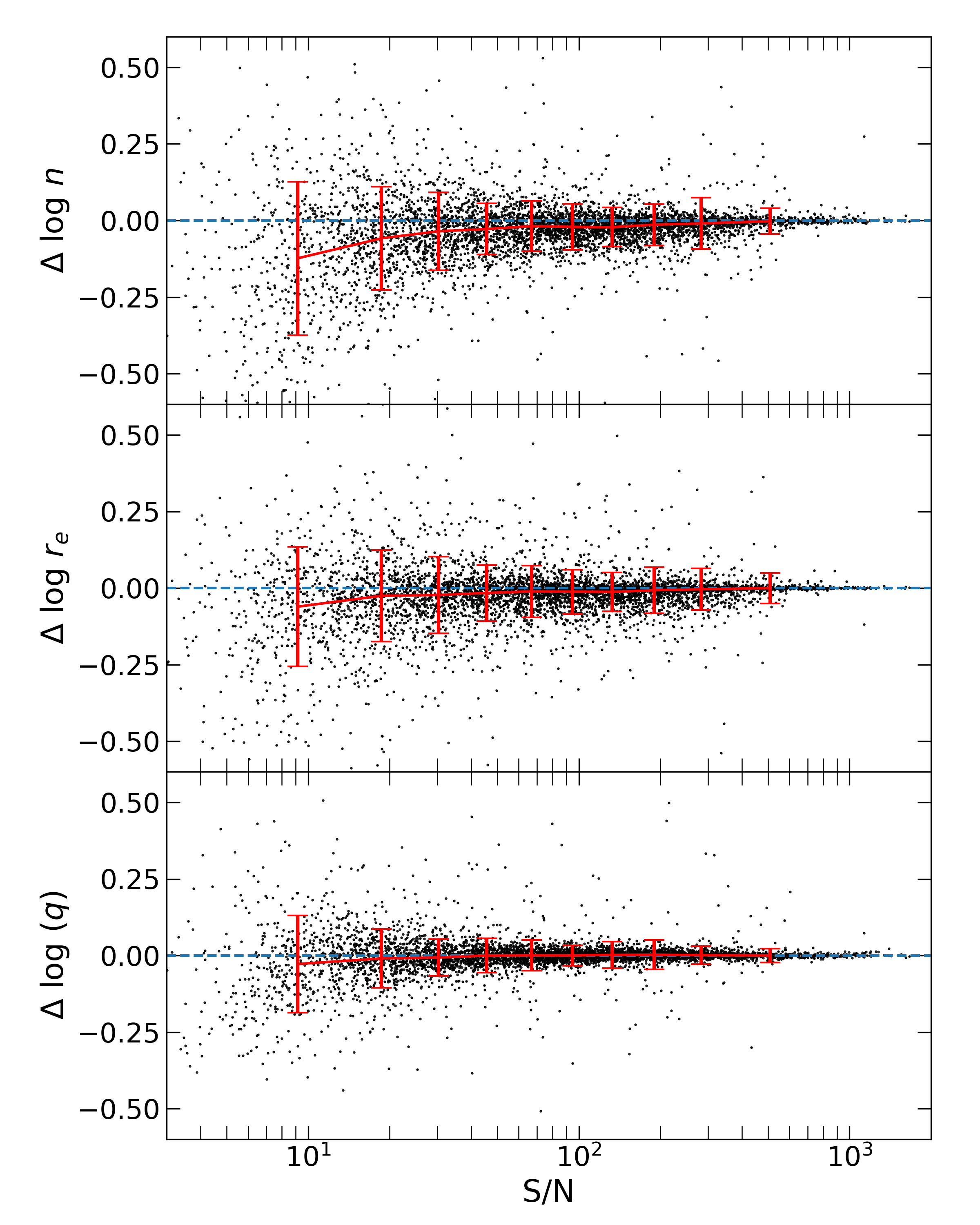}
    \caption{Difference between recovered and input parameters for simulated galaxies as a function of signal-to-noise ratio (SNR), where positive (negative) values represent larger (smaller) recovered parameters. Trends are shown in red for ten bins containing equal numbers of galaxies. Error bars represent the dispersion of values within each SNR bin. The blue dashed line represents equality between input and recovered parameters. Two features are evident: 1) at lower SNR, there is a larger dispersion, implying more uncertainty in the recovered parameters, and 2) at lower SNR we find that parameters are biased towards smaller recovered values. For S\'ersic index, $n$, we find the largest bias, reaching 0.12 dex at SNR=10, while $r_e$ is biased smaller by 0.05 dex and $q$ by 0.025 dex at the same SNR.}
    \label{fig:snr_bias}
\end{figure}

Returning to our simulated dataset, in figure \ref{fig:snr_bias} we show the difference between input parameters and recovered parameters for our simulated galaxies as a function of signal-to-noise ratio (SNR), where the SNR is computed from the aperture-corrected SExtractor photometry. Similar to \citet{vanderWel12}, we find that the scatter in recovered values (which corresponds to the overall uncertainty in any given parameter) increases at low SNR. In addition to this general trend, we find that there is a clear bias in the recovered values of $n$, $r_e$, and, to a lesser extent, $q$ as a function of SNR, with the recovered values becoming smaller as SNR decreases.  Our recovered $n$ are consistent with the input $n$ from the simulations at the highest SNR, while at SNR=10, the recovered $n$ is 0.12 dex smaller, on average.  Likewise, at the highest SNRs the recovered $r_e$ are consistent with the input $r_e$, but at SNR=10, the recovered $r_e$ is 0.05 dex smaller than the input $r_e$.  There is also some evidence that the recovered $q$ is smaller at the lowest SNR (0.025 dex at SNR=10), but there is very little bias over the majority of the SNR range. Thus, we expect lower-SNR data to be biased towards smaller $n$ and $r_e$. This makes intuitive sense given that errors in $n$ and $r_e$ are correlated (e.g., \S \ref{sec:errors-correlated}), and it is harder to detect the low-surface-brightness wings of high-$n$ sources with low-SNR data.  When compared with the CANDELS data, which has a lower SNR and poorer resolution than the CEERS data, this helps explain why our recovered S\'ersic indices are $14\%$ larger, and our sizes are $2\%$ larger, on average, over the same magnitude range.

\begin{figure*}
\label{fig:sizes_nir_vs_optical}
    \centering
    \includegraphics[scale=0.50]{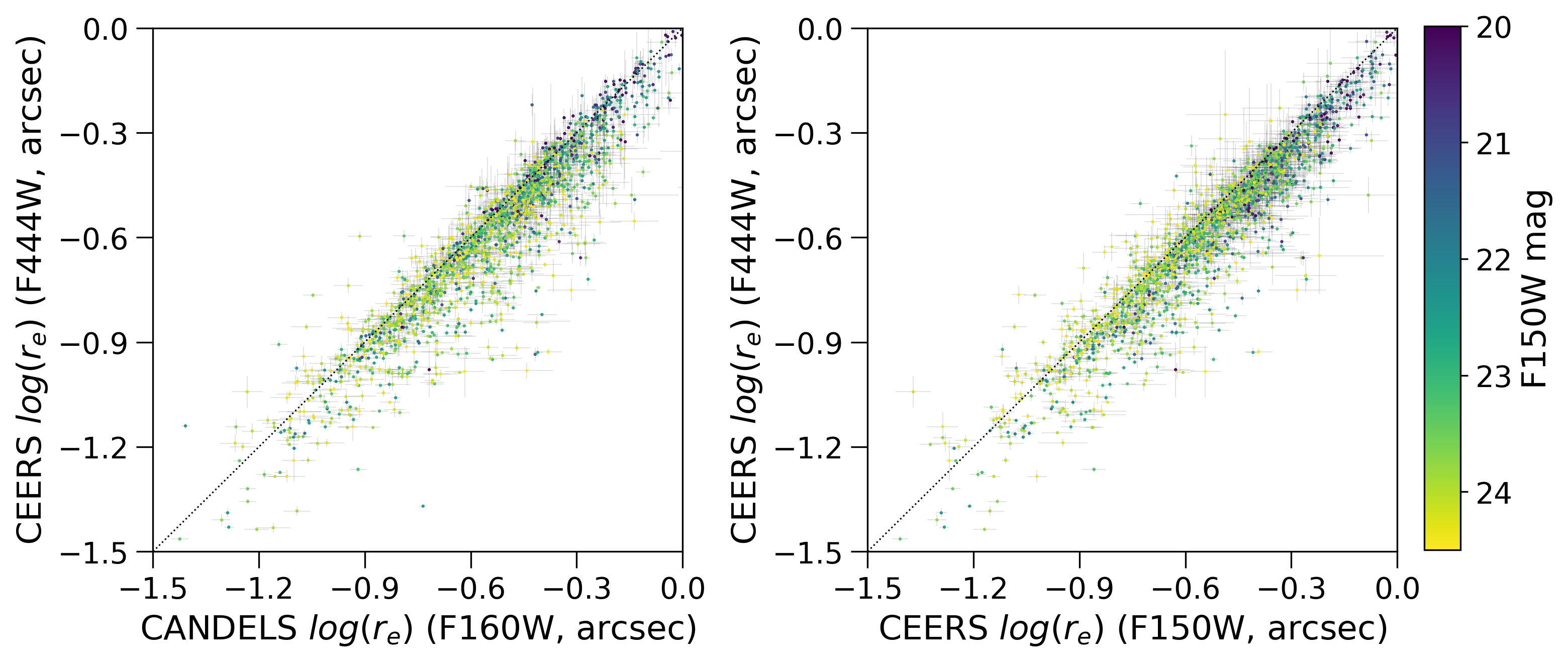}
    \caption{Comparison of galaxy sizes in short versus long wavelengths.Left: Sizes of galaxies detected in CANDELS imaging in the F160W filter versus sizes measured in CEERS at F444W. Right: Sizes for the same set of galaxies as in the left-hand panel, but measured entirely from CEERS imaging (F150W vs. F444W). Galaxies are smaller on average by 0.046 dex in F444W than in F150W or F160W, indicating that the rest-frame near-IR light, which more closely traces the mass distribution, is more centrally concentrated than the rest-frame optical light.}
\end{figure*}

Another important difference is that the effective radii measured in F444W with CEERS are systematically smaller than those measured in F160W with CANDELS by $11\%$ (Fig.~\ref{fig:sizes_nir_vs_optical}, left) for the same subset of galaxies (F160W$<24.5$~mag). The resolution of the CEERS F444W imaging (PSF FWHM = 0\farcs16) is comparable to the resolution of the CANDELS F160W imaging (PSF FWHM = 0\farcs17), suggesting that this is unlikely a resolution effect. We find a similar offset (0.046 dex) between the sizes of the exact same galaxies in the CEERS F150W and F444W images (Fig.~\ref{fig:sizes_nir_vs_optical}, right). This is consistent with findings by \citet{suess22} that measured galaxy sizes are smaller at rest-frame near-IR than at rest-frame optical, indicating that galaxy stellar mass is more concentrated than rest-frame optical light profiles suggest, as expected if galaxies follow an inside-out growth mechanism (e.g., \citealt{nelson16}). This in turn places important upper limits on the stellar mass content that these galaxies can host, which could have important implications at high-redshift, where stellar mass estimates are widely debated. However, it is also important to recognize that nebular emission can dominate the broadband flux density at certain wavelengths, especially at both lower masses and higher redshift (e.g., \citealt{smit14, davis24, llerena24}), and the extent or compactness of these nebular regions could have significant effects on the observed size of galaxies as a function of wavelength (e.g., \citealt{papaderos12}).
In section \ref{sec:color-gradients}, we explore the wavelength-dependence of galaxy size in more detail.

\subsection{Evolution in the Size-Mass Plane}
\label{sec:size-mass}
Sizes measurements have been used extensively in previous works to trace the assembly history of galaxies and their connection to dark matter halos (e.g., \citealt{shen03, vanderWel14, lange15, casura22, ormerod24, ward24, martorano24, allen25, yang25}). Exploiting the depth, resolution, multi-wavelength imaging, and wide area coverage of CEERS allows us to revisit the size evolution of galaxies as a function of mass at the same rest-frame wavelength across a large range of cosmic time from $1<z<9$.  

\begin{figure*}
    \centering
    \includegraphics[scale=0.60]{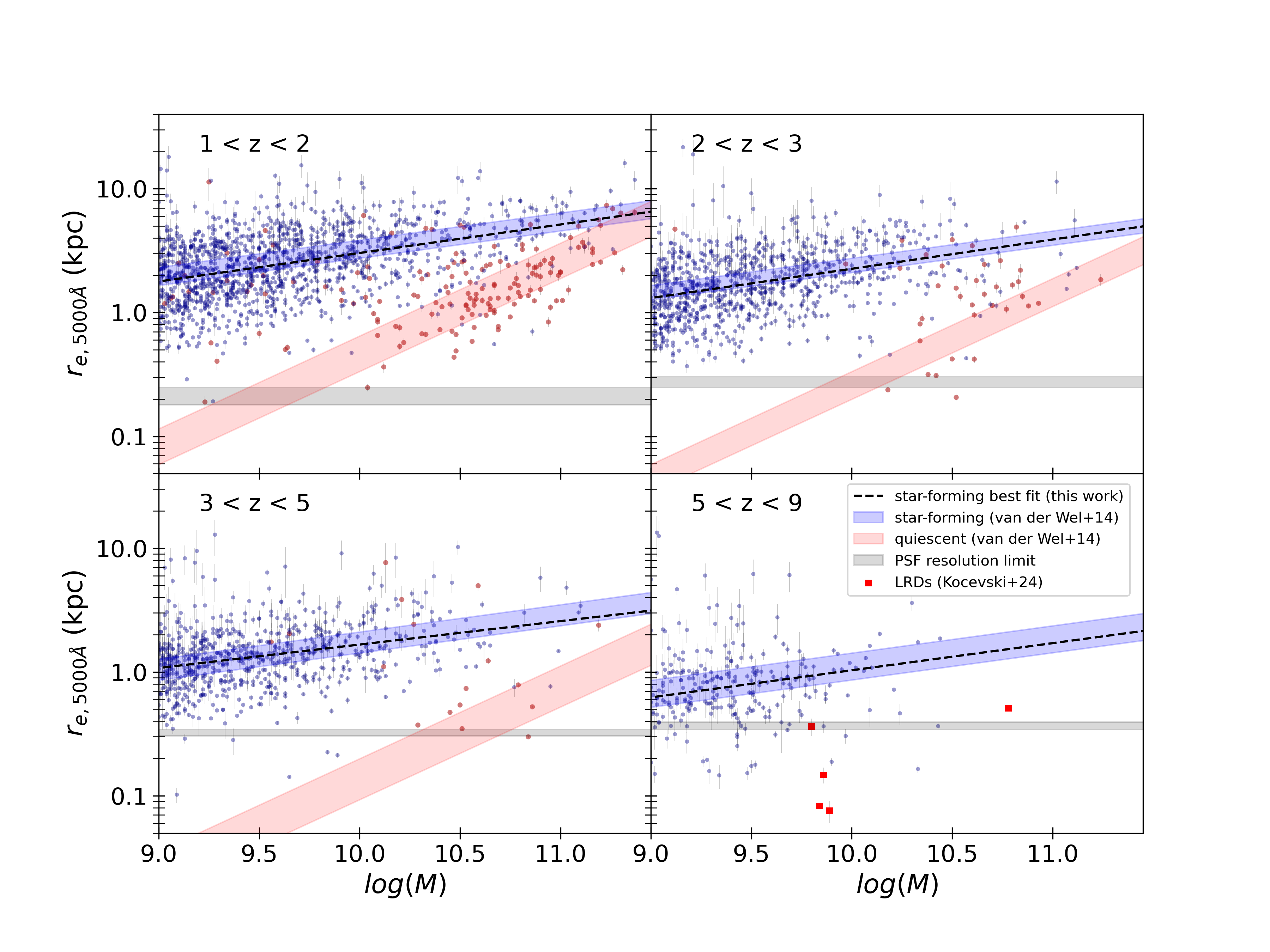}
    \caption{Galaxy size as a function of mass in four redshift bins. Data points are color-coded by their location in rest-frame UVJ color-color space (blue = star-forming, red = quiescent). Shaded regions show the best-fit relations from \citet{vanderWel14} extrapolated to each redshift range for star-forming and quiescent galaxies (blue and red, respectively). The dashed blue line in each plot is the best fit linear relation for star-forming galaxies using the CEERS data. Some of the most compact sources at high-z are identified as ``little red dots" (LRDs, from \citealt{Kocevski_2024}) and are excluded from the fit. The PSF resolution limit is shown as the horizontal gray shaded region for each redshift range.}
    \label{fig:size-mass}
\end{figure*}

In figure~\ref{fig:size-mass}, we show the rest-frame $5000$\AA~size as a function of mass for 3401 sources with $log(M) > 9.0$, flag values $<2$ and size errors $<50\%$ in four different redshift bins between $1<z<9$. Sizes, flags, and errors are determined from the filter that most closely matches rest-frame $5000$\AA~given the source redshift. The points are color-coded by their location in UVJ color-color space (e.g., \citealt{williams09}), with star-forming galaxies shown as blue and quiescent galaxies shown as red.  There are no quiescent galaxies in the highest redshift bin ($5<z<9$).  Overplotted on the CEERS data are the best fits from \citet{vanderWel14} for CANDELS data, extrapolated to these higher redshifts (shaded blue and red regions for star-forming and quiescent galaxies, respectively, with the extent of the shaded region covering the redshift range shown in each panel). 

At $z>2$, the number of quiescent galaxies in CEERS is too small to robustly fit the size-mass relation independently from star-forming galaxies, therefore, we fit a relation for the star-forming galaxies only, excluding any quiescent galaxies from the fit (dashed blue lines in fig.~\ref{fig:size-mass}).  We use the parameterization 

\begin{equation}
    \frac{R_e}{\rm{kpc}} = A\left( \frac{M_*}{5\times10^{10}M_{\odot}}\right)^{\alpha} 
\end{equation}
where $R_e$ is the effective radius at rest-frame $5000$~\AA~ and $M_*$ is the stellar mass from our SED fitting. Since we plot log$(R_e)$ against log$(M_*)$, $\alpha$ is the slope of the relation, and log$(A)$ is the intercept at $5\times10^{10}M_{\odot}$. Our results are given in table \ref{tab:size-mass-results}. We find that the slope, $\alpha$, is consistent over all three of the lowest redshift bins ($1<z<5$). Given the small number of galaxies in the highest redshift bin (227), we choose to hold the slope fixed at the average value from the lower three redshift bins. We also exclude 5 galaxies identified as ``little red dots'' (LRDs) in \citet{Kocevski_2024} since both their sizes and masses may be inaccurate. We find excellent agreement with the extrapolated results from \citet{vanderWel14}, with both an average slope, $\alpha=0.22$, and intercepts, log$(A)$, that are entirely consistent with \citet{vanderWel14} out to the highest redshifts.

\begin{deluxetable}{ccc}
    \tablewidth{3.5in}
    \label{tab:size-mass-results}
    \centering
    \tablecaption{Parameterized Fits to the Size-Mass Relation of Late-Type Galaxies (Shown in Figure~\ref{fig:size-mass})$^a$}
    \tablehead{
%         \colhead{$z$} & \hspace{0.3in} \colhead{log$(A)$}  &  \hspace{0.3in} \colhead{$\alpha$}}
         \colhead{$z$} &  \colhead{log$(A)$}  & \colhead{$\alpha$}}
    \startdata
        1.5    & \hspace{0.3in} $0.64\pm0.01$   &  \hspace{0.3in}  $0.23\pm0.01$ \\
        2.5    &  \hspace{0.3in} $0.52\pm0.01$  &  \hspace{0.3in}  $0.24\pm0.02$ \\
        4.0    &  \hspace{0.3in} $0.35\pm0.02$  &  \hspace{0.3in}  $0.19\pm0.02$ \\
        7.0    &  \hspace{0.3in} $0.17\pm0.02$  &  \hspace{0.3in}  $0.22\pm0.03^b$ \\
    \enddata
    \tablenotetext{a}{Assumes $R_e$/kpc = $A(M_*/5\times10^{10}M_{\odot})^{\alpha}$}
    \tablenotetext{b}{Average from the lowest three redshift bins.}
\end{deluxetable}

\subsection{Color Gradients}
\label{sec:color-gradients}
With S\'ersic profiles for each galaxy fit to the entire wavelength range covered by the CEERS NIRCam imaging, it is possible to look at the rest-frame optical to rest-frame near-infrared (NIR) color gradients of galaxies spanning  masses $\log(M)>10^9M_{\odot}$ and redshifts $z<5$. Color gradients hold clues to the evolutionary history of galaxies, by revealing differences in stellar populations that retain information imprinted on them by different formation scenarios and quenching processes. 

\begin{figure*}
    \centering
    \includegraphics[scale=0.70]{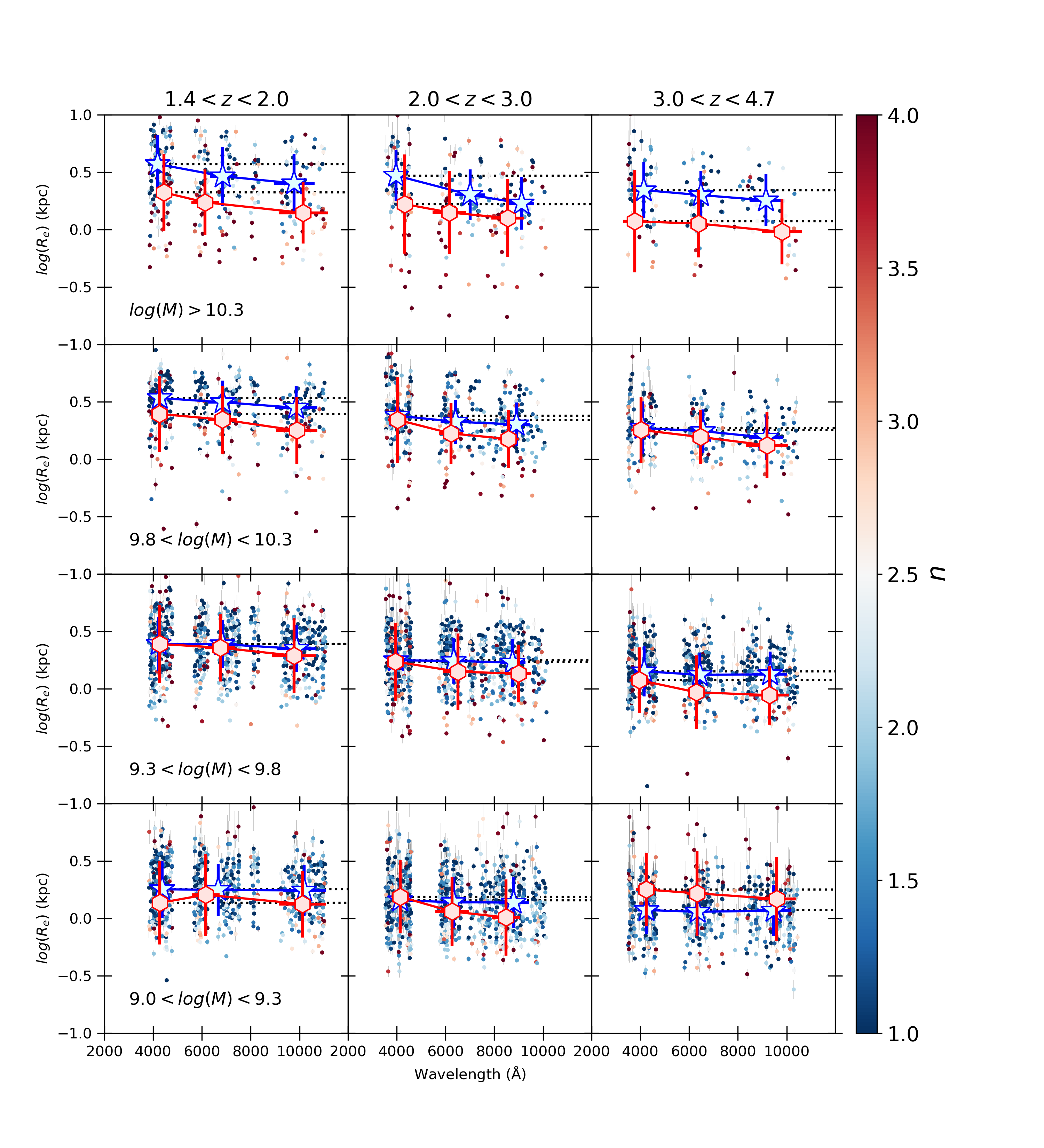}
    \caption{Galaxy size is shown as a function of rest-frame wavelength in four different mass bins and three different redshift bins. Individual points are color-coded by their S\'ersic index, while the average values for $n<2.5$ and $n>2.5$ are shown, along with their 1 sigma spread, as blue stars and red circles, respectively.  Dotted lines indicate the average size at the shortest wavelength, for reference. There is a clear trend that for the most massive galaxies, sizes are smaller at longer wavelengths for all redshifts probed. The trend is slightly less apparent in the highest-redshift bin for the highest-mass sources, but the number statistics are lower here, especially for $n>2.5$ sources. The trend of smaller galaxy sizes at longer wavelengths breaks down somewhere between $9.3 < log(M) < 9.8$ and there is no indication that galaxies with $log(M)<9.3$ show any variation in size as a function of wavelength.}
    \label{fig:color-gradients-mass-z}
\end{figure*}

Previous studies have noted that galaxies tend to be more compact at rest-frame NIR wavelengths \citep{suess22, allen25}, indicating that there are strong age, stellar metallicity, and/or attenuation gradients, especially among the most massive galaxies ($M_*\sim10^{11}M_{\odot}$), and that the mass profiles (which are better traced by NIR light) are more compact than rest-frame optical and UV light profiles suggest. To look at this in more detail, we define three approximate rest-frame wavelength regimes ($4000$\AA, $6500$\AA, and $9000$\AA) over the redshift range $1.4\leq z \leq 4.7$ and use the NIRCam broadband filters most closely matching those rest-wavelengths to measure half-light radii for galaxies with $\log(M_*)>10^9M_{\odot}$. The results are shown in figure \ref{fig:color-gradients-mass-z} for both high-S\'ersic (red) and low-S\'ersic (blue) galaxies in four different mass bins and three different redshift bins. The trend with mass, whereby the most massive galaxies exhibit the largest difference in half-light radii from rest-frame optical to rest-frame NIR, as noted by \citet{suess22}, is clearly evident for galaxies with both high- and low-S\'ersic indices. At masses $\log(M)<9.3M_{\odot}$, there is no discernible difference in galaxy size as a function of wavelength, while at masses $9.3<\log(M)<9.8$, only the high-$n$ galaxies show any perceptible difference in size as a function of wavelength.  Galaxies with high $n$ are smaller on average than galaxies with low $n$ at all wavelengths, evident at all masses and redshifts, which is consistent with the idea that quenched galaxies (which tend to have higher $n$) are smaller than star-forming galaxies at the same epoch (as shown in figure \ref{fig:size-mass}). 

Variations in size as a function of wavelength, however, could arise not only due to stellar population gradients, but also intrinsic differences in the morphology of a galaxy as a function of wavelength (e.g., the prominence of a bulge versus disk). To account for these structural differences, we can use the full S\'ersic profiles to calculate the flux density at some fiducial radius in different wavelengths in order to measure color as a function of radius. We follow a similar method as \citet{vanderWel24} and integrate the S\'ersic profiles to two common radii, 0.5$R_{e,0.65}$ and 2.0$R_{e,0.65}$, where $R_{e,0.65}$ is the effective semi-major axis measured in the NIRCam filter that most closely aligns with rest-frame $0.65\mu m$. Following \citet{vanderWel24}, we define the color gradient, measured using rest-frame $0.4\mu m$~and $0.9\mu m$~fluxes as 

\begin{equation}
    \label{eqn:color-gradient}
    \Delta C = \log \left(\frac{S_{0.9}(2R_{e,0.65}) / S_{0.4}(2R_{e,0.65})}{S_{0.9}(0.5R_{e,0.65}) / S_{0.4}(0.5R_{e,0.65})} \right)
\end{equation}

where $S_{0.9}(2R_{e,0.65})$ is the flux density at $0.9\mu m$ calculated by integrating the S\'ersic profile from the NIRCam filter most-closely matching rest-frame $0.9\mu m$ to $2R_{e,0.65}$. Negative values of the color gradient correspond to bluer outer regions and redder inner regions.

\begin{figure*}
    \centering
    \includegraphics[scale=0.70]{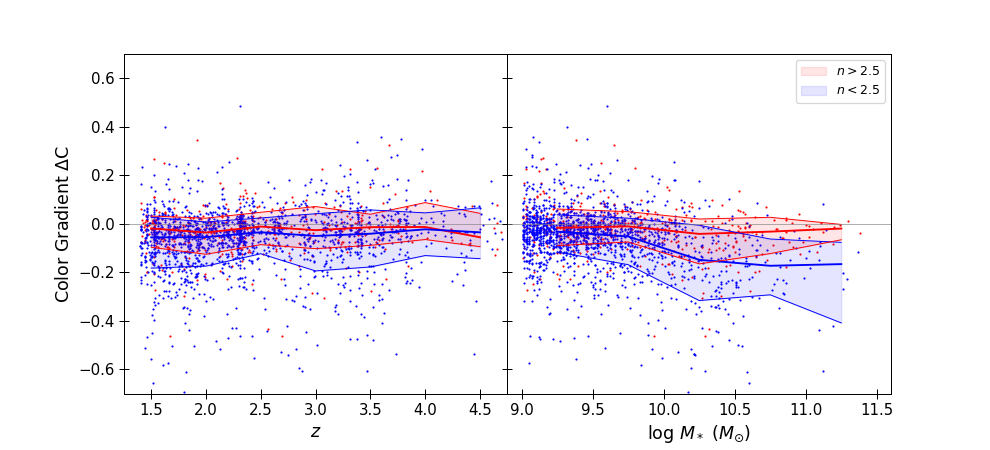}
    \caption{Color gradient, as defined in equation \ref{eqn:color-gradient}, is shown as a function of redshift and mass. Individual galaxy points are color-coded according to their S\'ersic index with $n<2.5$ shown as blue, and $n>2.5$ shown as red. The mean relation, along with the $16-84^{\rm{th}}$ percentile are shown for $n<2.5$ and $n>2.5$ galaxies as the blue and red lines and shaded regions, respectively. We find that there is no evidence for variation in color gradients as a function of redshift for either low- or high-S\'ersic galaxies, and color gradients are slightly negative on average. We also find evidence for larger color gradients among the most massive galaxies, a trend that is most evident for low-$n$ sources.}
    \label{fig:color-gradients}
\end{figure*}

In figure \ref{fig:color-gradients}, we show these color gradients as a function of redshift and mass. We split the sample into low- and high-$n$ samples (defined as $n<2.5$ and $n>2.5$, measured at rest-frame $0.65\mu m$, respectively).  The average color gradients are negative over all redshift and mass ranges, although the high-$n$ sample exhibits significantly flatter color gradients, particularly at the lowest masses (log($M_*)<10 M_{\odot}$). 

In this section we have separated galaxies based on their S\'ersic indices, as it is a key parameter provided in our morphology catalog and generally correlates well with other physical galaxy properties. However, it is important to note that extreme star-formers often have significant compact nebular emission that dominates the bulk of their rest-frame optical flux (e.g., \citealt{davis24}), and these galaxies may exhibit profiles more similar to small spheroidal systems (e.g., \citealt{bergvall02,caon05,amorin07,amorin09}). Thus, particularly at high$-z$ and lower masses, future work combining structural measurements with specific star-formation rates will be crucial to explore wavelength-dependent size variation and color gradients more broadly.

\section{Discussion} \label{sec:discussion}
Morphology measurements of a large sample of galaxies spanning a wide range in wavelength, with the resolution and depth of the JWST/CEERS imaging enables a wealth of scientific studies. We have demonstrated the utility of these measurements for two specific science cases: the evolution of galaxy size as a function of mass, and the evolution of color gradients as a function of mass.

Prior to JWST, evolution in the size-mass plane could only be explored to $z\sim3$ at rest-frame optical wavelengths, and only to intermediate masses ($\log(M_*)>9.5M_{\odot}$) at the highest redshifts. 
\citet{vanderWel14} found that for late-type galaxies, the slope of the mass-size relation is constant with redshift (to $z\sim3$) and follows the form $R_{e} \propto M^{\alpha}$, where $\alpha=0.22\pm0.03$. On the other hand, \citet{shen03} find that for local galaxies, there is a mass dependence on the slope, with more massive galaxies (log$(M)>10.6$) fit by a slope $\alpha=0.4$ and less massive galaxies fit with $\alpha=0.15$. Neither the \citet{vanderWel14} sample nor our own contains enough galaxies at log$(M)>10.6$ to explore this change in slope at high masses and at high-$z$, but we find an average slope of $\alpha=0.22\pm0.03$ over all redshifts which is consistent with \citet{vanderWel14} (see fig.~\ref{fig:size-mass}).  Similarly, there is evidence for a flattening of the mass-size relation for quiescent galaxies at the low-mass end (e.g., \citealt{cappellari13, lange15, nedkova21, kawinwanichakij21, cutler22, cutler25}). While our sample of quiescent galaxies is too small to address this, future work combining data from multiple wide-field JWST surveys will be important in studying both the low- and high-mass ends of the relation.

The redshift evolution of the intercept of the size-mass relation has typically been parameterized as a function of $(1+z)^{\beta_z}$, but \citet{vanderWel14} argue for a parameterization with $H(z)^{\beta_H}$ (with $\beta_H=-0.66$), which more closely aligns with the redshift evolution of the dark matter halo properties. For example, the evolution of the dark matter halo radius at fixed mass follows $R\propto H(z)^{-2/3}$ when defined with respect to the critical density, but follows a steeper relation when defined with respect to virial velocity, $R\propto H(z)^{-1}$ \citep{ferguson04}. Using this $H(z)^{\beta_H}$ parameterization, we find $r_e/$kpc = 7.6 h(z)$^{-0.66}$, which is in excellent agreement with \citet{vanderWel14}. The data do not allow for a significantly steeper evolution, such as $r_e\propto H(z)^{-1}$. 

If our sample is biased against detecting large, low-mass galaxies, this might affect both the slope and intercept of the size-mass relation, however \citet{pandya24} found (in their Appendix B) that our catalog is nearly 100\% complete for all sizes to $m_{\rm{F277W}}\leq26.5$.  Between $26.5<m_{\rm{F277W}}\leq27.0$ we start to become marginally incomplete to large, face-on disks, but these are at the extreme faint end of our sample, and most galaxies with log$(M)>9.0$ at $z<9$ are brighter than $m_{\rm{F277W}}=27.0$. 
Likewise, if our sample is biased against detecting small galaxies, this too could affect the resulting slope and intercept. In figure \ref{fig:size-mass}, we see that our measurements are well above the PSF resolution limit for all but the highest-redshift bin.  In the $5<z<9$ bin, it is possible that we miss some of the smallest galaxies, which could result in a lower intercept than we have determined (note that the slope is well-determined by the lower three redshift bins, with no evidence for any evolution).  However, even if our intercept in this bin is off by as much as 0.10~dex, we still cannot fit a $\beta_H<-0.75$. The pace of evolution is well-determined by the lower redshift bins, and ceases to be well-fit by either a $H(z)^{\beta_H}$ or a $(1+z)^{\beta_z}$ parameterization if the intercept in the highest-redshift bin is significantly lower than our determined value.
Thus, we find that our results are largely consistent with \citet{vanderWel14}. 
Likewise, our results are also consistent with recent results from JWST \citep{ormerod24, ward24, martorano24, allen25, yang25}.
Future work including larger samples of high$-z$ galaxies from other JWST survey fields will be important in confirming the rate of evolution at the earliest times, and in exploring the rate of evolution for early-type galaxies, which we do not explore here.

Color gradients have been even more challenging to study statistically, given the required spatial resolution and wavelength coverage needed to explore differences between the rest-frame UV, optical and NIR for galaxies beyond cosmic noon. 
\citet{szomoru11} analyzed the rest $u-g$ color gradients for a handful of galaxies out to $z\sim2$ and found that most $z\sim2$ galaxies have negative color gradients (redder centers), and that the average color gradient does not change with redshift. \citet{wuyts12} explored this with a larger sample of several hundred galaxies out to $z\sim2.5$, confirming the negative color gradients overall, and noting that the mass profiles are more centrally concentrated than the rest-frame optical light profiles, which is consistent with inside-out growth scenarios. 

More recently, \citet{vanderWel24} used CEERS JWST/NIRCam imaging to study the color gradients and NIR sizes of several hundred galaxies between $0.5<z<2.3$ to a mass limit of log$(M_*)>10.0$. Their color gradients are evaluated between rest-frame $4000$~\AA~ and $6000$~\AA.  They find that the average color gradients for both quiescent and star-forming galaxies are negative across the entire redshift range studied, and that the color gradients of star-forming galaxies exhibit a clear mass trend, whereby the most massive galaxies exhibit the most strongly negative color gradients. We find qualitatively similar results to \citet{vanderWel24} evaluating the color gradients between rest-frame $4000$~\AA~ and $9000$~\AA~ for galaxies between $1.4<z<4.7$ with log$(M_*)>9.0$. Even though our baseline wavelength range is larger than \citet{vanderWel24}, our average color gradients are smaller, likely because we probe to lower galaxy mass, where the color gradients are flatter. We also do not find quite as large a scatter in color gradient, likely because we do not have as many high-mass galaxies in our sample, especially at high$-z$.  Even so, we can confirm the trends found by \citet{vanderWel24}, including the finding that color gradients are generally negative and unchanging out to $z\sim5$ and that late-type galaxies have a strong mass dependence on their color gradient, with more massive galaxies exhibiting the largest color gradients overall.

If we compare our high-$n$ sample ($n>2.5$) from figures \ref{fig:color-gradients-mass-z} and \ref{fig:color-gradients}, we can see that while these galaxies are generally smaller at longer wavelengths, they do not exhibit particularly strong color gradients and the color gradients exhibit weak or no mass dependence. This indicates that the stellar populations of high-$n$ galaxies are largely uniform throughout, with likely only small negative age gradients from the center to the outermost regions, or possibly that the flat color gradients arise due to competing effects of age,  attenuation and metallicity gradients with opposing slopes.
Size differences for high-$n$ galaxies likely arise because of the prominence of a central concentration or bulge at different wavelengths.

On the other hand, low$-n$ galaxies show a wavelength dependence in size as a function of mass that is similar to the trend in color gradients with mass; the more massive the galaxy, the more significant the difference in size between rest-frame optical and NIR and the more strongly negative the color gradient. This indicates that wavelength-dependent size variation among the low$-n$ population may, in fact, be driven by substantial differences in the stellar populations and dust attenuation at different radii, or the extent of nebular regions, resulting in significant morphological $k$-corrections. 

Interpretation of wavelength-dependent size variation must also take into account differences in galaxy populations as a function of mass and redshift. For instance, extreme emission line galaxies (EELGs) which are common at $z>4$, and their low-mass counterparts at low-$z$, are often dominated by nebular emission from compact nuclear regions at rest-optical wavelengths. These galaxies may be best-fit by high S\'ersic indices, but their stellar populations may be more closely related to low-$n$ galaxies. Finally, the physical driver for the observed color gradients, particularly among massive galaxies, is still widely debated, with some simulations suggesting inside-out growth is the main driver (e.g., \citealt{baes24}), while others argue that dust attenuation alone can account for the size differences (e.g., \citealt{marshall22, nedkova24}). We therefore caution that further work on the wavelength-dependent size variation at both high$-z$ and lower masses will need larger samples that can accurately separate strongly star-forming from quiescent galaxies.

\section{Conclusions}
We provide the first public catalog of morphology for over 50,000 galaxies observed with JWST in order to enable community-led science that requires knowledge of galaxy size, shape, and structure up to $z\gtrsim10$.  Our measurements include:
\begin{itemize}
    \item Total magnitude, S\'ersic index, effective semi-major axis, axis ratio, and position angle for 53,885 galaxies observed with NIRCam in six filters (F115W, F150W, F200W, F277W, F356W, and F444W) in the CEERS public imaging.
    \item Quality flags representing the reliability of the fit, with FLAG=0 being good fits, FLAG=1 are lower quality fits, FLAG=2 are fits that reached a constraint limit and which should be used with caution, FLAG=3 are fits that failed, and FLAG=4 are suspected image artifacts.
    \item Robust uncertainties on the measured parameters based on the scatter in recovered values for a set of simulated galaxies spanning a similar distribution to CEERS galaxies and with similar noise characteristics.
\end{itemize}

We have further demonstrated that our results are consistent with previous HST-based studies of galaxy morphology \citep{vanderWel12} with lower SNR and resolution for a subset of the data where measurements were deemed reliable in both studies. As one approaches lower SNR, both the size and S\'ersic index are biased low, emphasizing that the choice to impose a magnitude or SNR-limit for use of these morphological parameters is appropriate for most studies.  We find that the majority of measurements of basic size and shape parameters have a 20\% accuracy or better at the following magnitudes: $m_{\rm{F115W}}\leq26.5$, $m_{\rm{F150W}}\leq26.25$, $m_{\rm{F200W}}\leq26.5$, $m_{\rm{F277W}}\leq26.75$, $m_{\rm{F356W}}\leq27.0$, and $m_{\rm{F444W}}\leq26.4$.

Further, choice of PSF does not appear to affect the overall resulting morphology by a significant amount, except for the highest-S\'ersic sources ($n>2.5$), where the different PSFs we tested yielded S\'ersic indices and sizes that differed by nearly 24\% and 10\%, respectively, or for the most compact sources ($r_e<0\farcs18$), where the different PSFs yielded S\'ersic indices and sizes that differed by roughly 17\% and 3\%, respectively.

Finally, we demonstrate the utility of these measurements by presenting the evolution in the size-mass plane of galaxies down to $log(M_*)>9$ out to $z<9$, as well as a look at how galaxy sizes depend on wavelength and how color gradients of early-type galaxies ($n<2.5$) depend strongly on mass. Future studies including morphology from other deep JWST public fields will help improve on the number statistics at the highest redshifts and help determine the precise way in which size evolves with mass in the early universe, thus constraining some of the most uncertain aspects of galaxy formation theory.

\begin{acknowledgments}
We thank the anonymous referee for helpful comments that improved the manuscript. FB acknowledges support from the GEELSBE2 project with reference PID2023-150393NB-I00 funded by MCIU/AEI/10.13039/501100011033 and the FSE+ and also the Consolidación Investigadora IGADLE project with reference CNS2024-154572. FB also acknowledges support from the project PID2020-116188GA-I00, funded by MICIU/AEI /10.13039/501100011033. FB gratefully acknowledges financial support of the Department of Education, Junta de Castilla y Le\'on and FEDER Funds (Reference: CLU-2023-1-05). The project that gave rise to these results received the support of a fellowship from the “la Caixa” Foundation (ID 100010434). The fellowship code is LCF/BQ/PR24/12050015. LC acknowledges support from grants PID2022-139567NB-I00 and PIB2021-127718NB-I00 funded by the Spanish Ministry of Science and Innovation/State Agency of Research  MCIN/AEI/10.13039/501100011033 and by “ERDF A way of making Europe”. PGP-G acknowledges support from grant PID2022-139567NB-I00 funded by Spanish Ministerio de Ciencia e Innovaci\'on MCIN/AEI/10.13039/501100011033,
FEDER {\it Una manera de hacer Europa}.

\end{acknowledgments}

\vspace{5mm}
\facilities{JWST(NIRCam)}

\software{{\sc galfit} \citep{peng10},  
          Source Extractor \citep{bertin96}
          }

\bibliography{galfit_ceers}{}

\begin{thebibliography}{}
\expandafter\ifx\csname natexlab\endcsname\relax\def\natexlab#1{#1}\fi
\providecommand{\url}[1]{\href{#1}{#1}}
\providecommand{\dodoi}[1]{doi:~\href{http://doi.org/#1}{\nolinkurl{#1}}}
\providecommand{\doeprint}[1]{\href{http://ascl.net/#1}{\nolinkurl{http://ascl.net/#1}}}
\providecommand{\doarXiv}[1]{\href{https://arxiv.org/abs/#1}{\nolinkurl{https://arxiv.org/abs/#1}}}

\bibitem[{{Allen} {et~al.}(2025){Allen}, {Oesch}, {Toft}, {Matharu},
  {McPartland}, {Weibel}, {Brammer}, {Bowler}, {Ito}, {Gottumukkala}, {Rizzo},
  {Valentino}, {Varadaraj}, {Weaver}, \& {Whitaker}}]{allen25}
{Allen}, N., {Oesch}, P.~A., {Toft}, S., {et~al.} 2025, \aap, 698, A30,
  \dodoi{10.1051/0004-6361/202452690}

\bibitem[{{Allen} {et~al.}(2017){Allen}, {Kacprzak}, {Glazebrook}, {Labb{\'e}},
  {Tran}, {Spitler}, {Cowley}, {Nanayakkara}, {Papovich}, {Quadri},
  {Straatman}, {Tilvi}, \& {van Dokkum}}]{allen17}
{Allen}, R.~J., {Kacprzak}, G.~G., {Glazebrook}, K., {et~al.} 2017, \apjl, 834,
  L11, \dodoi{10.3847/2041-8213/834/2/L11}

\bibitem[{{Amor{\'\i}n} {et~al.}(2009){Amor{\'\i}n}, {Aguerri},
  {Mu{\~n}oz-Tu{\~n}{\'o}n}, \& {Cair{\'o}s}}]{amorin09}
{Amor{\'\i}n}, R., {Aguerri}, J.~A.~L., {Mu{\~n}oz-Tu{\~n}{\'o}n}, C., \&
  {Cair{\'o}s}, L.~M. 2009, \aap, 501, 75, \dodoi{10.1051/0004-6361/200809591}

\bibitem[{{Amor{\'\i}n} {et~al.}(2007){Amor{\'\i}n}, {Mu{\~n}oz-Tu{\~n}{\'o}n},
  {Aguerri}, {Cair{\'o}s}, \& {Caon}}]{amorin07}
{Amor{\'\i}n}, R.~O., {Mu{\~n}oz-Tu{\~n}{\'o}n}, C., {Aguerri}, J.~A.~L.,
  {Cair{\'o}s}, L.~M., \& {Caon}, N. 2007, \aap, 467, 541,
  \dodoi{10.1051/0004-6361:20066152}

\bibitem[{{Baes} {et~al.}(2024){Baes}, {Mosenkov}, {Kelly}, {Abdurro'uf},
  {Andreadis}, {Bokona Tulu}, {Camps}, {Tassama Emana}, {Fritz}, {Gebek},
  {Kova{\v{c}}i{\'c}}, {La Marca}, {Martorano}, {Nersesian}, {Rodriguez-Gomez},
  {Tortora}, {Tr{\v{c}}ka}, {Vander Meulen}, {van der Wel}, \& {Wang}}]{baes24}
{Baes}, M., {Mosenkov}, A., {Kelly}, R., {et~al.} 2024, \aap, 683, A182,
  \dodoi{10.1051/0004-6361/202348419}

\bibitem[{{Baggen} {et~al.}(2023){Baggen}, {van Dokkum}, {Labb{\'e}},
  {Brammer}, {Miller}, {Bezanson}, {Leja}, {Wang}, {Whitaker}, {Suess}, \&
  {Nelson}}]{baggen23}
{Baggen}, J. F.~W., {van Dokkum}, P., {Labb{\'e}}, I., {et~al.} 2023, \apjl,
  955, L12, \dodoi{10.3847/2041-8213/acf5ef}

\bibitem[{{Bagley} {et~al.}(2023){Bagley}, {Finkelstein}, {Koekemoer},
  {Ferguson}, {Arrabal Haro}, {Dickinson}, {Kartaltepe}, {Papovich},
  {P{\'e}rez-Gonz{\'a}lez}, {Pirzkal}, {Somerville}, {Willmer}, {Yang}, {Yung},
  {Fontana}, {Grazian}, {Grogin}, {Hirschmann}, {Kewley}, {Kirkpatrick},
  {Kocevski}, {Lotz}, {Medrano}, {Morales}, {Pentericci}, {Ravindranath},
  {Trump}, {Wilkins}, {Calabr{\`o}}, {Cooper}, {Costantin}, {de la Vega},
  {Hilbert}, {Hutchison}, {Larson}, {Lucas}, {McGrath}, {Ryan}, {Wang}, \&
  {Wuyts}}]{bagley23}
{Bagley}, M.~B., {Finkelstein}, S.~L., {Koekemoer}, A.~M., {et~al.} 2023,
  \apjl, 946, L12, \dodoi{10.3847/2041-8213/acbb08}

\bibitem[{{Bergvall} \& {{\"O}stlin}(2002)}]{bergvall02}
{Bergvall}, N., \& {{\"O}stlin}, G. 2002, \aap, 390, 891,
  \dodoi{10.1051/0004-6361:20020759}

\bibitem[{{Bertin} \& {Arnouts}(1996)}]{bertin96}
{Bertin}, E., \& {Arnouts}, S. 1996, \aaps, 117, 393,
  \dodoi{10.1051/aas:1996164}

\bibitem[{{Bertin} {et~al.}(2020){Bertin}, {Schefer}, {Apostolakos},
  {{\'A}lvarez-Ayll{\'o}n}, {Dubath}, \& {K{\"u}mmel}}]{bertin20}
{Bertin}, E., {Schefer}, M., {Apostolakos}, N., {et~al.} 2020, in Astronomical
  Society of the Pacific Conference Series, Vol. 527, Astronomical Data
  Analysis Software and Systems XXIX, ed. R.~{Pizzo}, E.~R. {Deul}, J.~D.
  {Mol}, J.~{de Plaa}, \& H.~{Verkouter}, 461

\bibitem[{{Brammer} {et~al.}(2008){Brammer}, {van Dokkum}, \&
  {Coppi}}]{brammer08}
{Brammer}, G.~B., {van Dokkum}, P.~G., \& {Coppi}, P. 2008, \apj, 686, 1503,
  \dodoi{10.1086/591786}

\bibitem[{{Brammer} {et~al.}(2012){Brammer}, {van Dokkum}, {Franx},
  {Fumagalli}, {Patel}, {Rix}, {Skelton}, {Kriek}, {Nelson}, {Schmidt},
  {Bezanson}, {da Cunha}, {Erb}, {Fan}, {F{\"o}rster Schreiber}, {Illingworth},
  {Labb{\'e}}, {Leja}, {Lundgren}, {Magee}, {Marchesini}, {McCarthy},
  {Momcheva}, {Muzzin}, {Quadri}, {Steidel}, {Tal}, {Wake}, {Whitaker}, \&
  {Williams}}]{brammer12}
{Brammer}, G.~B., {van Dokkum}, P.~G., {Franx}, M., {et~al.} 2012, \apjs, 200,
  13, \dodoi{10.1088/0067-0049/200/2/13}

\bibitem[{{Bruzual} \& {Charlot}(2003)}]{bc03}
{Bruzual}, G., \& {Charlot}, S. 2003, \mnras, 344, 1000,
  \dodoi{10.1046/j.1365-8711.2003.06897.x}

\bibitem[{{Buitrago} {et~al.}(2008){Buitrago}, {Trujillo}, {Conselice},
  {Bouwens}, {Dickinson}, \& {Yan}}]{buitrago08}
{Buitrago}, F., {Trujillo}, I., {Conselice}, C.~J., {et~al.} 2008, \apjl, 687,
  L61, \dodoi{10.1086/592836}

\bibitem[{{Calzetti} {et~al.}(2000){Calzetti}, {Armus}, {Bohlin}, {Kinney},
  {Koornneef}, \& {Storchi-Bergmann}}]{calzetti00}
{Calzetti}, D., {Armus}, L., {Bohlin}, R.~C., {et~al.} 2000, \apj, 533, 682,
  \dodoi{10.1086/308692}

\bibitem[{{Caon} {et~al.}(2005){Caon}, {Cair{\'o}s}, {Aguerri}, \&
  {Mu{\~n}oz-Tu{\~n}{\'o}n}}]{caon05}
{Caon}, N., {Cair{\'o}s}, L.~M., {Aguerri}, J. A.~L., \&
  {Mu{\~n}oz-Tu{\~n}{\'o}n}, C. 2005, \apjs, 157, 218, \dodoi{10.1086/428286}

\bibitem[{{Cappellari} {et~al.}(2013){Cappellari}, {McDermid}, {Alatalo},
  {Blitz}, {Bois}, {Bournaud}, {Bureau}, {Crocker}, {Davies}, {Davis}, {de
  Zeeuw}, {Duc}, {Emsellem}, {Khochfar}, {Krajnovi{\'c}}, {Kuntschner},
  {Morganti}, {Naab}, {Oosterloo}, {Sarzi}, {Scott}, {Serra}, {Weijmans}, \&
  {Young}}]{cappellari13}
{Cappellari}, M., {McDermid}, R.~M., {Alatalo}, K., {et~al.} 2013, \mnras, 432,
  1862, \dodoi{10.1093/mnras/stt644}

\bibitem[{{Cassata} {et~al.}(2013){Cassata}, {Giavalisco}, {Williams}, {Guo},
  {Lee}, {Renzini}, {Ferguson}, {Faber}, {Barro}, {McIntosh}, {Lu}, {Bell},
  {Koo}, {Papovich}, {Ryan}, {Conselice}, {Grogin}, {Koekemoer}, \&
  {Hathi}}]{cassata13}
{Cassata}, P., {Giavalisco}, M., {Williams}, C.~C., {et~al.} 2013, \apj, 775,
  106, \dodoi{10.1088/0004-637X/775/2/106}

\bibitem[{{Casura} {et~al.}(2022){Casura}, {Liske}, {Robotham}, {Brough},
  {Driver}, {Graham}, {H{\"a}u{\ss}ler}, {Holwerda}, {Hopkins}, {Kelvin},
  {Moffett}, {Taranu}, \& {Taylor}}]{casura22}
{Casura}, S., {Liske}, J., {Robotham}, A. S.~G., {et~al.} 2022, \mnras, 516,
  942, \dodoi{10.1093/mnras/stac2267}

\bibitem[{{Chabrier}(2003)}]{chabrier03}
{Chabrier}, G. 2003, Publications of the Astronomical Society of the Pacific,
  115, 763, \dodoi{10.1086/376392}

\bibitem[{{Conroy} {et~al.}(2010){Conroy}, {White}, \& {Gunn}}]{conroy10}
{Conroy}, C., {White}, M., \& {Gunn}, J.~E. 2010, \apj, 708, 58,
  \dodoi{10.1088/0004-637X/708/1/58}

\bibitem[{{Costantin} {et~al.}(2023){Costantin}, {P{\'e}rez-Gonz{\'a}lez},
  {Guo}, {Buttitta}, {Jogee}, {Bagley}, {Barro}, {Kartaltepe}, {Koekemoer},
  {Cabello}, {Corsini}, {M{\'e}ndez-Abreu}, {de la Vega}, {Iyer}, {Bisigello},
  {Cheng}, {Morelli}, {Arrabal Haro}, {Buitrago}, {Cooper}, {Dekel},
  {Dickinson}, {Finkelstein}, {Giavalisco}, {Holwerda}, {Huertas-Company},
  {Lucas}, {Papovich}, {Pirzkal}, {Seill{\'e}}, {Vega-Ferrero}, {Wuyts}, \&
  {Yung}}]{costantin23}
{Costantin}, L., {P{\'e}rez-Gonz{\'a}lez}, P.~G., {Guo}, Y., {et~al.} 2023,
  \nat, 623, 499, \dodoi{10.1038/s41586-023-06636-x}

\bibitem[{{Costantin} {et~al.}(2025){Costantin}, {Gillman}, {Boogaard},
  {P{\'e}rez-Gonz{\'a}lez}, {Iani}, {Rinaldi}, {Melinder}, {Crespo G{\'o}mez},
  {Colina}, {Greve}, {{\"O}stlin}, {Wright}, {Alonso-Herrero},
  {{\'A}lvarez-M{\'a}rquez}, {Annunziatella}, {Bik}, {Caputi}, {Dicken},
  {Eckart}, {Hjorth}, {Ilbert}, {Jermann}, {Labiano}, {Langeroodi},
  {Pei{\ss}ker}, {Pye}, {Tikkanen}, {van der Werf}, {Walter}, {Ward},
  {G{\"u}del}, \& {Henning}}]{costantin25}
{Costantin}, L., {Gillman}, S., {Boogaard}, L.~A., {et~al.} 2025, \aap, 699,
  A360, \dodoi{10.1051/0004-6361/202451330}

\bibitem[{{Cutler} {et~al.}(2022){Cutler}, {Whitaker}, {Mowla}, {Brammer}, {van
  der Wel}, {Marchesini}, {van Dokkum}, {Momcheva}, {Song}, {Akhshik},
  {Nelson}, {Bezanson}, {Franx}, {Kriek}, {Lange-Vagle}, {Leja}, {MacKenty},
  {Muzzin}, \& {Shipley}}]{cutler22}
{Cutler}, S.~E., {Whitaker}, K.~E., {Mowla}, L.~A., {et~al.} 2022, \apj, 925,
  34, \dodoi{10.3847/1538-4357/ac341c}

\bibitem[{{Cutler} {et~al.}(2025){Cutler}, {Weaver}, {Whitaker}, {Greene},
  {Setton}, {Webb}, {Abdullah}, {Medrano}, {Bezanson}, {Brammer}, {Feldmann},
  {Furtak}, {Glazebrook}, {Labbe}, {Leja}, {Marchesini}, {Miller},
  {Mitsuhashi}, {Nanayakkara}, {Nelson}, {Pan}, {Price}, {Suess}, \&
  {Wang}}]{cutler25}
{Cutler}, S.~E., {Weaver}, J.~R., {Whitaker}, K.~E., {et~al.} 2025, \apj, 993,
  169, \dodoi{10.3847/1538-4357/ae0629}

\bibitem[{{Davis} {et~al.}(2024){Davis}, {Trump}, {Simons}, {McGrath},
  {Wilkins}, {Arrabal Haro}, {Bagley}, {Dickinson}, {Fern{\'a}ndez},
  {Amor{\'\i}n}, {Backhaus}, {Cleri}, {Llerena}, {Brunker}, {Barro},
  {Bisigello}, {Brooks}, {Costantin}, {de La Vega}, {Dekel}, {Finkelstein},
  {Hathi}, {Hirschmann}, {Kartaltepe}, {Koekemoer}, {Lucas}, {Papovich},
  {P{\'e}rez-Gonz{\'a}lez}, {Pirzkal}, {Rodighiero}, {Rose}, {Yung}, \& {Ceers
  Collaborators}}]{davis24}
{Davis}, K., {Trump}, J.~R., {Simons}, R.~C., {et~al.} 2024, \apj, 974, 42,
  \dodoi{10.3847/1538-4357/ad6865}

\bibitem[{{Davis} {et~al.}(2007){Davis}, {Guhathakurta}, {Konidaris}, {Newman},
  {Ashby}, {Biggs}, {Barmby}, {Bundy}, {Chapman}, {Coil}, {Conselice},
  {Cooper}, {Croton}, {Eisenhardt}, {Ellis}, {Faber}, {Fang}, {Fazio},
  {Georgakakis}, {Gerke}, {Goss}, {Gwyn}, {Harker}, {Hopkins}, {Huang},
  {Ivison}, {Kassin}, {Kirby}, {Koekemoer}, {Koo}, {Laird}, {Le Floc'h}, {Lin},
  {Lotz}, {Marshall}, {Martin}, {Metevier}, {Moustakas}, {Nandra}, {Noeske},
  {Papovich}, {Phillips}, {Rich}, {Rieke}, {Rigopoulou}, {Salim},
  {Schiminovich}, {Simard}, {Smail}, {Small}, {Weiner}, {Willmer}, {Willner},
  {Wilson}, {Wright}, \& {Yan}}]{davis07}
{Davis}, M., {Guhathakurta}, P., {Konidaris}, N.~P., {et~al.} 2007, \apjl, 660,
  L1, \dodoi{10.1086/517931}

\bibitem[{{de Vaucouleurs}(1961)}]{devauc61}
{de Vaucouleurs}, G. 1961, \apjs, 5, 233, \dodoi{10.1086/190056}

\bibitem[{{Ferguson} {et~al.}(2004){Ferguson}, {Dickinson}, {Giavalisco},
  {Kretchmer}, {Ravindranath}, {Idzi}, {Taylor}, {Conselice}, {Fall},
  {Gardner}, {Livio}, {Madau}, {Moustakas}, {Papovich}, {Somerville},
  {Spinrad}, \& {Stern}}]{ferguson04}
{Ferguson}, H.~C., {Dickinson}, M., {Giavalisco}, M., {et~al.} 2004, \apjl,
  600, L107, \dodoi{10.1086/378578}

\bibitem[{{Ferreira} {et~al.}(2023){Ferreira}, {Conselice}, {Sazonova},
  {Ferrari}, {Caruana}, {Tohill}, {Lucatelli}, {Adams}, {Irodotou}, {Marshall},
  {Roper}, {Lovell}, {Verma}, {Austin}, {Trussler}, \& {Wilkins}}]{ferreira23}
{Ferreira}, L., {Conselice}, C.~J., {Sazonova}, E., {et~al.} 2023, \apj, 955,
  94, \dodoi{10.3847/1538-4357/acec76}

\bibitem[{{Finkelstein} {et~al.}(2022){Finkelstein}, {Bagley}, {Arrabal Haro},
  {Dickinson}, {Ferguson}, {Kartaltepe}, {Papovich}, {Burgarella}, {Kocevski},
  {Huertas-Company}, {Iyer}, {Koekemoer}, {Larson}, {P{\'e}rez-Gonz{\'a}lez},
  {Rose}, {Tacchella}, {Wilkins}, {Chworowsky}, {Medrano}, {Morales},
  {Somerville}, {Yung}, {Fontana}, {Giavalisco}, {Grazian}, {Grogin}, {Kewley},
  {Kirkpatrick}, {Kurczynski}, {Lotz}, {Pentericci}, {Pirzkal}, {Ravindranath},
  {Ryan}, {Trump}, {Yang}, {Almaini}, {Amor{\'\i}n}, {Annunziatella},
  {Backhaus}, {Barro}, {Behroozi}, {Bell}, {Bhatawdekar}, {Bisigello}, {Bromm},
  {Buat}, {Buitrago}, {Calabr{\`o}}, {Casey}, {Castellano}, {Ch{\'a}vez Ortiz},
  {Ciesla}, {Cleri}, {Cohen}, {Cole}, {Cooke}, {Cooper}, {Cooray}, {Costantin},
  {Cox}, {Croton}, {Daddi}, {Dav{\'e}}, {de La Vega}, {Dekel}, {Elbaz},
  {Estrada-Carpenter}, {Faber}, {Fern{\'a}ndez}, {Finkelstein}, {Freundlich},
  {Fujimoto}, {Garc{\'\i}a-Argum{\'a}nez}, {Gardner}, {Gawiser},
  {G{\'o}mez-Guijarro}, {Guo}, {Hamblin}, {Hamilton}, {Hathi}, {Holwerda},
  {Hirschmann}, {Hutchison}, {Jaskot}, {Jha}, {Jogee}, {Juneau}, {Jung},
  {Kassin}, {Le Bail}, {Leung}, {Lucas}, {Magnelli}, {Mantha}, {Matharu},
  {McGrath}, {McIntosh}, {Merlin}, {Mobasher}, {Newman}, {Nicholls}, {Pandya},
  {Rafelski}, {Ronayne}, {Santini}, {Seill{\'e}}, {Shah}, {Shen}, {Simons},
  {Snyder}, {Stanway}, {Straughn}, {Teplitz}, {Vanderhoof}, {Vega-Ferrero},
  {Wang}, {Weiner}, {Willmer}, {Wuyts}, {Zavala}, \& {Ceers
  Team}}]{finkelstein22}
{Finkelstein}, S.~L., {Bagley}, M.~B., {Arrabal Haro}, P., {et~al.} 2022,
  \apjl, 940, L55, \dodoi{10.3847/2041-8213/ac966e}

\bibitem[{{Finkelstein} {et~al.}(2023){Finkelstein}, {Bagley}, {Ferguson},
  {Wilkins}, {Kartaltepe}, {Papovich}, {Yung}, {Arrabal Haro}, {Behroozi},
  {Dickinson}, {Kocevski}, {Koekemoer}, {Larson}, {Le Bail}, {Morales},
  {P{\'e}rez-Gonz{\'a}lez}, {Burgarella}, {Dav{\'e}}, {Hirschmann},
  {Somerville}, {Wuyts}, {Bromm}, {Casey}, {Fontana}, {Fujimoto}, {Gardner},
  {Giavalisco}, {Grazian}, {Grogin}, {Hathi}, {Hutchison}, {Jha}, {Jogee},
  {Kewley}, {Kirkpatrick}, {Long}, {Lotz}, {Pentericci}, {Pierel}, {Pirzkal},
  {Ravindranath}, {Ryan}, {Trump}, {Yang}, {Bhatawdekar}, {Bisigello}, {Buat},
  {Calabr{\`o}}, {Castellano}, {Cleri}, {Cooper}, {Croton}, {Daddi}, {Dekel},
  {Elbaz}, {Franco}, {Gawiser}, {Holwerda}, {Huertas-Company}, {Jaskot},
  {Leung}, {Lucas}, {Mobasher}, {Pandya}, {Tacchella}, {Weiner}, \&
  {Zavala}}]{finkelstein23}
{Finkelstein}, S.~L., {Bagley}, M.~B., {Ferguson}, H.~C., {et~al.} 2023, \apjl,
  946, L13, \dodoi{10.3847/2041-8213/acade4}

\bibitem[{{Finkelstein} {et~al.}(2025){Finkelstein}, {Bagley}, {Arrabal Haro},
  {Dickinson}, {Ferguson}, {Kartaltepe}, {Kocevski}, {Koekemoer}, {Lotz},
  {Papovich}, {P{\'e}rez-Gonz{\'a}lez}, {Pirzkal}, {Somerville}, {Trump},
  {Yang}, {Yung}, {Fontana}, {Grazian}, {Grogin}, {Kewley}, {Kirkpatrick},
  {Larson}, {Pentericci}, {Ravindranath}, {Wilkins}, {Almaini}, {Amor{\'\i}n},
  {Barro}, {Bhatawdekar}, {Bisigello}, {Brooks}, {Buat}, {Buitrago},
  {Burgarella}, {Calabr{\`o}}, {Castellano}, {Cheng}, {Cleri}, {Cole},
  {Cooper}, {Cooper}, {Costantin}, {Cox}, {Croton}, {Daddi}, {Davis}, {Dekel},
  {Elbaz}, {Fern{\'a}ndez}, {Fujimoto}, {Gandolfi}, {Gardner}, {Gawiser},
  {Giavalisco}, {G{\'o}mez-Guijarro}, {Guo}, {Gupta}, {Hathi}, {Harish},
  {Henry}, {Hirschmann}, {Hu}, {Hutchison}, {Iyer}, {Jaskot}, {Jha}, {Jung},
  {Kassin}, {Kokorev}, {Kurczynski}, {Leung}, {Llerena}, {Long}, {Lucas}, {Lu},
  {McGrath}, {McIntosh}, {Merlin}, {Mobasher}, {Morales}, {Napolitano},
  {Pacucci}, {Pandya}, {Rafelski}, {Rodighiero}, {Rose}, {Santini},
  {Seill{\'e}}, {Simons}, {Shen}, {Straughn}, {Tacchella}, {Taylor},
  {Vanderhoof}, {Vega-Ferrero}, {Weiner}, {Willmer}, {Zhu}, {Bell}, {Wuyts},
  {Holwerda}, {Wang}, {Wang}, {Zavala}, \& {CEERS Collaboration}}]{ceers25}
{Finkelstein}, S.~L., {Bagley}, M.~B., {Arrabal Haro}, P., {et~al.} 2025,
  \apjl, 983, L4, \dodoi{10.3847/2041-8213/adbbd3}

\bibitem[{{Gillman} {et~al.}(2024){Gillman}, {Smail}, {Gullberg}, {Swinbank},
  {Vijayan}, {Lee}, {Brammer}, {Dudzevi{\v{c}}i{\={u}}t{\.{e}}}, {Greve},
  {Almaini}, {Brinch}, {Chapman}, {Chen}, {Ikarashi}, {Matsuda}, {Wang},
  {Walter}, \& {van der Werf}}]{gillman24}
{Gillman}, S., {Smail}, I., {Gullberg}, B., {et~al.} 2024, \aap, 691, A299,
  \dodoi{10.1051/0004-6361/202451006}

\bibitem[{{Graham} {et~al.}(1996){Graham}, {Lauer}, {Colless}, \&
  {Postman}}]{graham96}
{Graham}, A., {Lauer}, T.~R., {Colless}, M., \& {Postman}, M. 1996, \apj, 465,
  534, \dodoi{10.1086/177440}

\bibitem[{{Grogin} {et~al.}(2011){Grogin}, {Kocevski}, {Faber}, {Ferguson},
  {Koekemoer}, {Riess}, {Acquaviva}, {Alexander}, {Almaini}, {Ashby}, {Barden},
  {Bell}, {Bournaud}, {Brown}, {Caputi}, {Casertano}, {Cassata}, {Challis},
  {Chary}, {Cheung}, {Cirasuolo}, {Conselice}, {Roshan Cooray}, {Croton},
  {Daddi}, {Dahlen}, {Dav{\'e}}, {de Mello}, {Dekel}, {Dickinson}, {Dolch},
  {Donley}, {Dunlop}, {Dutton}, {Elbaz}, {Fazio}, {Filippenko}, {Finkelstein},
  {Fontana}, {Gardner}, {Garnavich}, {Gawiser}, {Giavalisco}, {Grazian}, {Guo},
  {Hathi}, {H{\"a}ussler}, {Hopkins}, {Huang}, {Huang}, {Jha}, {Kartaltepe},
  {Kirshner}, {Koo}, {Lai}, {Lee}, {Li}, {Lotz}, {Lucas}, {Madau}, {McCarthy},
  {McGrath}, {McIntosh}, {McLure}, {Mobasher}, {Moustakas}, {Mozena}, {Nandra},
  {Newman}, {Niemi}, {Noeske}, {Papovich}, {Pentericci}, {Pope}, {Primack},
  {Rajan}, {Ravindranath}, {Reddy}, {Renzini}, {Rix}, {Robaina}, {Rodney},
  {Rosario}, {Rosati}, {Salimbeni}, {Scarlata}, {Siana}, {Simard}, {Smidt},
  {Somerville}, {Spinrad}, {Straughn}, {Strolger}, {Telford}, {Teplitz},
  {Trump}, {van der Wel}, {Villforth}, {Wechsler}, {Weiner}, {Wiklind}, {Wild},
  {Wilson}, {Wuyts}, {Yan}, \& {Yun}}]{grogin11}
{Grogin}, N.~A., {Kocevski}, D.~D., {Faber}, S.~M., {et~al.} 2011, \apjs, 197,
  35.
\newblock \doarXiv{1105.3753}

\bibitem[{{Guo} {et~al.}(2009){Guo}, {McIntosh}, {Mo}, {Katz}, {Van Den Bosch},
  {Weinberg}, {Weinmann}, {Pasquali}, \& {Yang}}]{guo09}
{Guo}, Y., {McIntosh}, D.~H., {Mo}, H.~J., {et~al.} 2009, \mnras, 398, 1129,
  \dodoi{10.1111/j.1365-2966.2009.15223.x}

\bibitem[{{H{\"a}ussler} {et~al.}(2007){H{\"a}ussler}, {McIntosh}, {Barden},
  {Bell}, {Rix}, {Borch}, {Beckwith}, {Caldwell}, {Heymans}, {Jahnke}, {Jogee},
  {Koposov}, {Meisenheimer}, {S{\'a}nchez}, {Somerville}, {Wisotzki}, \&
  {Wolf}}]{haussler07}
{H{\"a}ussler}, B., {McIntosh}, D.~H., {Barden}, M., {et~al.} 2007, \apjs, 172,
  615, \dodoi{10.1086/518836}

\bibitem[{{Holmberg}(1958)}]{holmberg58}
{Holmberg}, E. 1958, Meddelanden fran Lunds Astronomiska Observatorium Serie
  II, 136, 1

\bibitem[{{Hubble}(1936)}]{hubble36}
{Hubble}, E.~P. 1936, {Realm of the Nebulae}

\bibitem[{{Huertas-Company} {et~al.}(2023){Huertas-Company}, {Iyer},
  {Angeloudi}, {Bagley}, {Finkelstein}, {Kartaltepe}, {Sarmiento},
  {Vega-Ferrero}, {Arrabal Haro}, {Behroozi}, {Buitrago}, {Cheng}, {Costantin},
  {Dekel}, {Dickinson}, {Elbaz}, {Grogin}, {Hathi}, {Holwerda}, {Koekemoer},
  {Lucas}, {Papovich}, {P{\'e}rez-Gonz{\'a}lez}, {Pirzkal}, {Seill{\'e}}, {de
  la Vega}, {Wuyts}, {Yang}, \& {Yung}}]{huertas-company23}
{Huertas-Company}, M., {Iyer}, K.~G., {Angeloudi}, E., {et~al.} 2023, arXiv
  e-prints, arXiv:2305.02478, \dodoi{10.48550/arXiv.2305.02478}

\bibitem[{{Jacobs} {et~al.}(2023){Jacobs}, {Glazebrook}, {Calabr{\`o}}, {Treu},
  {Nannayakkara}, {Jones}, {Merlin}, {Abraham}, {Stevens}, {Vulcani}, {Yang},
  {Bonchi}, {Boyett}, {Brada{\v{c}}}, {Castellano}, {Fontana}, {Marchesini},
  {Malkan}, {Mason}, {Morishita}, {Paris}, {Santini}, {Trenti}, \&
  {Wang}}]{jacobs23}
{Jacobs}, C., {Glazebrook}, K., {Calabr{\`o}}, A., {et~al.} 2023, \apjl, 948,
  L13, \dodoi{10.3847/2041-8213/accd6d}

\bibitem[{{Kartaltepe} {et~al.}(2023){Kartaltepe}, {Rose}, {Vanderhoof},
  {McGrath}, {Costantin}, {Cox}, {Yung}, {Kocevski}, {Wuyts}, {Ferguson},
  {Bagley}, {Finkelstein}, {Amor{\'\i}n}, {Andrews}, {Arrabal Haro},
  {Backhaus}, {Behroozi}, {Bisigello}, {Calabr{\`o}}, {Casey}, {Coogan},
  {Cooper}, {Croton}, {de la Vega}, {Dickinson}, {Fontana}, {Franco},
  {Grazian}, {Grogin}, {Hathi}, {Holwerda}, {Huertas-Company}, {Iyer}, {Jogee},
  {Jung}, {Kewley}, {Kirkpatrick}, {Koekemoer}, {Liu}, {Lotz}, {Lucas},
  {Newman}, {Pacifici}, {Pandya}, {Papovich}, {Pentericci},
  {P{\'e}rez-Gonz{\'a}lez}, {Petersen}, {Pirzkal}, {Rafelski}, {Ravindranath},
  {Simons}, {Snyder}, {Somerville}, {Stanway}, {Straughn}, {Tacchella},
  {Trump}, {Vega-Ferrero}, {Wilkins}, {Yang}, \& {Zavala}}]{kartaltepe23}
{Kartaltepe}, J.~S., {Rose}, C., {Vanderhoof}, B.~N., {et~al.} 2023, \apjl,
  946, L15, \dodoi{10.3847/2041-8213/acad01}

\bibitem[{{Kawinwanichakij} {et~al.}(2021){Kawinwanichakij}, {Silverman},
  {Ding}, {George}, {Damjanov}, {Sawicki}, {Tanaka}, {Taranu}, {Birrer},
  {Huang}, {Li}, {Onodera}, {Shibuya}, \& {Yasuda}}]{kawinwanichakij21}
{Kawinwanichakij}, L., {Silverman}, J.~D., {Ding}, X., {et~al.} 2021, \apj,
  921, 38, \dodoi{10.3847/1538-4357/ac1f21}

\bibitem[{{Kocevski} {et~al.}(2024){Kocevski}, {Finkelstein}, {Barro},
  {Taylor}, {Calabr{\`o}}, {Laloux}, {Buchner}, {Trump}, {Leung}, {Yang},
  {Dickinson}, {P{\'e}rez-Gonz{\'a}lez}, {Pacucci}, {Inayoshi}, {Somerville},
  {McGrath}, {Akins}, {Bagley}, {Bisigello}, {Bowler}, {Carnall}, {Casey},
  {Cheng}, {Cleri}, {Costantin}, {Cullen}, {Davis}, {Donnan}, {Dunlop},
  {Ellis}, {Ferguson}, {Fujimoto}, {Fontana}, {Giavalisco}, {Grazian},
  {Grogin}, {Hathi}, {Hirschmann}, {Huertas-Company}, {Holwerda},
  {Illingworth}, {Juneau}, {Kartaltepe}, {Koekemoer}, {Li}, {Lucas}, {Magee},
  {Mason}, {McLeod}, {McLure}, {Napolitano}, {Papovich}, {Pirzkal},
  {Rodighiero}, {Santini}, {Wilkins}, \& {Yung}}]{Kocevski_2024}
{Kocevski}, D.~D., {Finkelstein}, S.~L., {Barro}, G., {et~al.} 2024, arXiv
  e-prints, arXiv:2404.03576, \dodoi{10.48550/arXiv.2404.03576}

\bibitem[{{Koekemoer} {et~al.}(2011){Koekemoer}, {Faber}, {Ferguson}, {Grogin},
  {Kocevski}, {Koo}, {Lai}, {Lotz}, {Lucas}, {McGrath}, {Ogaz}, {Rajan},
  {Riess}, {Rodney}, {Strolger}, {Casertano}, {Dahlen}, {Dickinson}, {Dolch},
  {Fontana}, {Giavalisco}, {Grazian}, {Guo}, {Hathi}, {Huang}, {van der Wel},
  {Yan}, {Acquaviva}, {Almaini}, {Ashby}, {Barden}, {Bell}, {Bournaud},
  {Brown}, {Caputi}, {Cassata}, {Challis}, {Chary}, {Cheung}, {Cirasuolo},
  {Conselice}, {Roshan Cooray}, {Croton}, {Daddi}, {Dav{\'e}}, {de Mello}, {de
  Ravel}, {Dekel}, {Donley}, {Dunlop}, {Dutton}, {Elbaz}, {Fazio},
  {Filippenko}, {Finkelstein}, {Frazer}, {Gardner}, {Garnavich}, {Gawiser},
  {Gruetzbauch}, {Hartley}, {H{\"a}ussler}, {Herrington}, {Hopkins}, {Huang},
  {Jha}, {Johnson}, {Kartaltepe}, {Khostovan}, {Kirshner}, {Lani}, {Lee}, {Li},
  {Madau}, {McCarthy}, {McIntosh}, {McLure}, {McPartland}, {Mobasher},
  {Moreira}, {Mortlock}, {Moustakas}, {Mozena}, {Nandra}, {Newman}, {Nielsen},
  {Niemi}, {Noeske}, {Papovich}, {Pentericci}, {Pope}, {Primack},
  {Ravindranath}, {Reddy}, {Renzini}, {Rix}, {Robaina}, {Rosario}, {Rosati},
  {Salimbeni}, {Scarlata}, {Siana}, {Simard}, {Smidt}, {Snyder}, {Somerville},
  {Spinrad}, {Straughn}, {Telford}, {Teplitz}, {Trump}, {Vargas}, {Villforth},
  {Wagner}, {Wandro}, {Wechsler}, {Weiner}, {Wiklind}, {Wild}, {Wilson},
  {Wuyts}, \& {Yun}}]{koekemoer11}
{Koekemoer}, A.~M., {Faber}, S.~M., {Ferguson}, H.~C., {et~al.} 2011, \apjs,
  197, 36.
\newblock \doarXiv{1105.3754}

\bibitem[{{Kriek} {et~al.}(2009){Kriek}, {van Dokkum}, {Labb{\'e}}, {Franx},
  {Illingworth}, {Marchesini}, \& {Quadri}}]{kriek09}
{Kriek}, M., {van Dokkum}, P.~G., {Labb{\'e}}, I., {et~al.} 2009, \apj, 700,
  221, \dodoi{10.1088/0004-637X/700/1/221}

\bibitem[{{K{\"u}mmel} {et~al.}(2022){K{\"u}mmel}, {{\'A}lvarez-Ayll{\'o}n},
  {Bertin}, {Dubath}, {Gavazzi}, {Hartley}, \& {Schefer}}]{kummel22}
{K{\"u}mmel}, M., {{\'A}lvarez-Ayll{\'o}n}, A., {Bertin}, E., {et~al.} 2022,
  arXiv e-prints, arXiv:2212.02428, \dodoi{10.48550/arXiv.2212.02428}

\bibitem[{{Lange} {et~al.}(2015){Lange}, {Driver}, {Robotham}, {Kelvin},
  {Graham}, {Alpaslan}, {Andrews}, {Baldry}, {Bamford}, {Bland-Hawthorn},
  {Brough}, {Cluver}, {Conselice}, {Davies}, {Haeussler}, {Konstantopoulos},
  {Loveday}, {Moffett}, {Norberg}, {Phillipps}, {Taylor},
  {L{\'o}pez-S{\'a}nchez}, \& {Wilkins}}]{lange15}
{Lange}, R., {Driver}, S.~P., {Robotham}, A. S.~G., {et~al.} 2015, \mnras, 447,
  2603, \dodoi{10.1093/mnras/stu2467}

\bibitem[{{Lange} {et~al.}(2016){Lange}, {Moffett}, {Driver}, {Robotham},
  {Lagos}, {Kelvin}, {Conselice}, {Margalef-Bentabol}, {Alpaslan}, {Baldry},
  {Bland-Hawthorn}, {Bremer}, {Brough}, {Cluver}, {Colless}, {Davies},
  {H{\"a}u{\ss}ler}, {Holwerda}, {Hopkins}, {Kafle}, {Kennedy}, {Liske},
  {Phillipps}, {Popescu}, {Taylor}, {Tuffs}, {van Kampen}, \&
  {Wright}}]{lange16}
{Lange}, R., {Moffett}, A.~J., {Driver}, S.~P., {et~al.} 2016, \mnras, 462,
  1470, \dodoi{10.1093/mnras/stw1495}

\bibitem[{{Larson} {et~al.}(2023){Larson}, {Hutchison}, {Bagley},
  {Finkelstein}, {Yung}, {Somerville}, {Hirschmann}, {Brammer}, {Holwerda},
  {Papovich}, {Morales}, \& {Wilkins}}]{larson23}
{Larson}, R.~L., {Hutchison}, T.~A., {Bagley}, M., {et~al.} 2023, \apj, 958,
  141, \dodoi{10.3847/1538-4357/acfed4}

\bibitem[{{Lee} {et~al.}(2013){Lee}, {Giavalisco}, {Williams}, {Guo}, {Lotz},
  {Van der Wel}, {Ferguson}, {Faber}, {Koekemoer}, {Grogin}, {Kocevski},
  {Conselice}, {Wuyts}, {Dekel}, {Kartaltepe}, \& {Bell}}]{lee13}
{Lee}, B., {Giavalisco}, M., {Williams}, C.~C., {et~al.} 2013, \apj, 774, 47,
  \dodoi{10.1088/0004-637X/774/1/47}

\bibitem[{{Llerena} {et~al.}(2024){Llerena}, {Amor{\'\i}n}, {Pentericci},
  {Arrabal Haro}, {Backhaus}, {Bagley}, {Calabr{\`o}}, {Cleri}, {Davis},
  {Dickinson}, {Finkelstein}, {Gawiser}, {Grogin}, {Hathi}, {Hirschmann},
  {Kartaltepe}, {Koekemoer}, {McGrath}, {Mobasher}, {Napolitano}, {Papovich},
  {Pirzkal}, {Trump}, {Wilkins}, \& {Yung}}]{llerena24}
{Llerena}, M., {Amor{\'\i}n}, R., {Pentericci}, L., {et~al.} 2024, \aap, 691,
  A59, \dodoi{10.1051/0004-6361/202449904}

\bibitem[{{Madau} \& {Dickinson}(2014)}]{madau14}
{Madau}, P., \& {Dickinson}, M. 2014, \araa, 52, 415,
  \dodoi{10.1146/annurev-astro-081811-125615}

\bibitem[{{Marshall} {et~al.}(2022){Marshall}, {Wilkins}, {Di Matteo}, {Roper},
  {Vijayan}, {Ni}, {Feng}, \& {Croft}}]{marshall22}
{Marshall}, M.~A., {Wilkins}, S., {Di Matteo}, T., {et~al.} 2022, \mnras, 511,
  5475, \dodoi{10.1093/mnras/stac380}

\bibitem[{{Martorano} {et~al.}(2024){Martorano}, {van der Wel}, {Baes}, {Bell},
  {Brammer}, {Franx}, \& {Nersesian}}]{martorano24}
{Martorano}, M., {van der Wel}, A., {Baes}, M., {et~al.} 2024, \apj, 972, 134,
  \dodoi{10.3847/1538-4357/ad5c6a}

\bibitem[{{Mosleh} {et~al.}(2012){Mosleh}, {Williams}, {Franx}, {Gonzalez},
  {Bouwens}, {Oesch}, {Labbe}, {Illingworth}, \& {Trenti}}]{mosleh12}
{Mosleh}, M., {Williams}, R.~J., {Franx}, M., {et~al.} 2012, \apjl, 756, L12,
  \dodoi{10.1088/2041-8205/756/1/L12}

\bibitem[{{Mowla} {et~al.}(2019){Mowla}, {van Dokkum}, {Brammer}, {Momcheva},
  {van der Wel}, {Whitaker}, {Nelson}, {Bezanson}, {Muzzin}, {Franx},
  {MacKenty}, {Leja}, {Kriek}, \& {Marchesini}}]{mowla19}
{Mowla}, L.~A., {van Dokkum}, P., {Brammer}, G.~B., {et~al.} 2019, \apj, 880,
  57, \dodoi{10.3847/1538-4357/ab290a}

\bibitem[{{Nedkova} {et~al.}(2021){Nedkova}, {H{\"a}u{\ss}ler}, {Marchesini},
  {Dimauro}, {Brammer}, {Eigenthaler}, {Feinstein}, {Ferguson},
  {Huertas-Company}, {Johnston}, {Kado-Fong}, {Kartaltepe}, {Labb{\'e}},
  {Lange-Vagle}, {Martis}, {McGrath}, {Muzzin}, {Oesch}, {Ordenes-Brice{\~n}o},
  {Puzia}, {Shipley}, {Simmons}, {Skelton}, {Stefanon}, {van der Wel}, \&
  {Whitaker}}]{nedkova21}
{Nedkova}, K.~V., {H{\"a}u{\ss}ler}, B., {Marchesini}, D., {et~al.} 2021,
  \mnras, 506, 928, \dodoi{10.1093/mnras/stab1744}

\bibitem[{{Nedkova} {et~al.}(2024){Nedkova}, {Rafelski}, {Teplitz}, {Mehta},
  {Degroot}, {Ravindranath}, {Alavi}, {Beckett}, {Grogin}, {H{\"a}u{\ss}ler},
  {Koekemoer}, {Oyarz{\'u}n}, {Prichard}, {Revalski}, {Snyder}, {Sunnquist},
  {Wang}, {Windhorst}, {Chartab}, {Conselice}, {Guo}, {Hathi}, {Hayes}, {Ji},
  {Kim}, {Lucas}, {Mobasher}, {O'Connell}, {Sattari}, {Smith}, {Taamoli},
  {Yung}, \& {The Uvcandels Team}}]{nedkova24}
{Nedkova}, K.~V., {Rafelski}, M., {Teplitz}, H.~I., {et~al.} 2024, \apj, 970,
  188, \dodoi{10.3847/1538-4357/ad4ede}

\bibitem[{{Nelson} {et~al.}(2016){Nelson}, {van Dokkum}, {F{\"o}rster
  Schreiber}, {Franx}, {Brammer}, {Momcheva}, {Wuyts}, {Whitaker}, {Skelton},
  {Fumagalli}, {Hayward}, {Kriek}, {Labb{\'e}}, {Leja}, {Rix}, {Tacconi}, {van
  der Wel}, {van den Bosch}, {Oesch}, {Dickey}, \& {Ulf Lange}}]{nelson16}
{Nelson}, E.~J., {van Dokkum}, P.~G., {F{\"o}rster Schreiber}, N.~M., {et~al.}
  2016, \apj, 828, 27, \dodoi{10.3847/0004-637X/828/1/27}

\bibitem[{{Nelson} {et~al.}(2023){Nelson}, {Suess}, {Bezanson}, {Price}, {van
  Dokkum}, {Leja}, {Wang}, {Whitaker}, {Labb{\'e}}, {Barrufet}, {Brammer},
  {Eisenstein}, {Gibson}, {Hartley}, {Johnson}, {Heintz}, {Mathews}, {Miller},
  {Oesch}, {Sandles}, {Setton}, {Speagle}, {Tacchella}, {Tadaki}, {{\"U}bler},
  \& {Weaver}}]{nelson23}
{Nelson}, E.~J., {Suess}, K.~A., {Bezanson}, R., {et~al.} 2023, \apjl, 948,
  L18, \dodoi{10.3847/2041-8213/acc1e1}

\bibitem[{{Ono} {et~al.}(2013){Ono}, {Ouchi}, {Curtis-Lake}, {Schenker},
  {Ellis}, {McLure}, {Dunlop}, {Robertson}, {Koekemoer}, {Bowler}, {Rogers},
  {Schneider}, {Charlot}, {Stark}, {Shimasaku}, {Furlanetto}, \&
  {Cirasuolo}}]{ono13}
{Ono}, Y., {Ouchi}, M., {Curtis-Lake}, E., {et~al.} 2013, \apj, 777, 155,
  \dodoi{10.1088/0004-637X/777/2/155}

\bibitem[{{Ormerod} {et~al.}(2024){Ormerod}, {Conselice}, {Adams}, {Harvey},
  {Austin}, {Trussler}, {Ferreira}, {Caruana}, {Lucatelli}, {Li}, \&
  {Roper}}]{ormerod24}
{Ormerod}, K., {Conselice}, C.~J., {Adams}, N.~J., {et~al.} 2024, \mnras, 527,
  6110, \dodoi{10.1093/mnras/stad3597}

\bibitem[{{Pandya} {et~al.}(2024){Pandya}, {Zhang}, {Huertas-Company}, {Iyer},
  {McGrath}, {Barro}, {Finkelstein}, {K{\"u}mmel}, {Hartley}, {Ferguson},
  {Kartaltepe}, {Primack}, {Dekel}, {Faber}, {Koo}, {Bryan}, {Somerville},
  {Amor{\'\i}n}, {Arrabal Haro}, {Bagley}, {Bell}, {Bertin}, {Costantin},
  {Dav{\'e}}, {Dickinson}, {Feldmann}, {Fontana}, {Gavazzi}, {Giavalisco},
  {Grazian}, {Grogin}, {Guo}, {Hahn}, {Holwerda}, {Kewley}, {Kirkpatrick},
  {Kocevski}, {Koekemoer}, {Lotz}, {Lucas}, {Papovich}, {Pentericci},
  {P{\'e}rez-Gonz{\'a}lez}, {Pirzkal}, {Ravindranath}, {Rose}, {Schefer},
  {Simons}, {Straughn}, {Tacchella}, {Trump}, {de la Vega}, {Wilkins}, {Wuyts},
  {Yang}, \& {Yung}}]{pandya24}
{Pandya}, V., {Zhang}, H., {Huertas-Company}, M., {et~al.} 2024, \apj, 963, 54,
  \dodoi{10.3847/1538-4357/ad1a13}

\bibitem[{{Pandya} {et~al.}(2025){Pandya}, {Loeb}, {McGrath}, {Barro},
  {Finkelstein}, {Ferguson}, {Grogin}, {Kartaltepe}, {Koekemoer}, {Papovich},
  {Pirzkal}, \& {Yung}}]{pandya24b}
{Pandya}, V., {Loeb}, A., {McGrath}, E.~J., {et~al.} 2025, \apj, 986, 72,
  \dodoi{10.3847/1538-4357/adcd78}

\bibitem[{{Papaderos} \& {{\"O}stlin}(2012)}]{papaderos12}
{Papaderos}, P., \& {{\"O}stlin}, G. 2012, \aap, 537, A126,
  \dodoi{10.1051/0004-6361/201117551}

\bibitem[{{Papovich} {et~al.}(2005){Papovich}, {Dickinson}, {Giavalisco},
  {Conselice}, \& {Ferguson}}]{papovich05}
{Papovich}, C., {Dickinson}, M., {Giavalisco}, M., {Conselice}, C.~J., \&
  {Ferguson}, H.~C. 2005, \apj, 631, 101, \dodoi{10.1086/429120}

\bibitem[{{Peng} {et~al.}(2010){Peng}, {Ho}, {Impey}, \& {Rix}}]{peng10}
{Peng}, C.~Y., {Ho}, L.~C., {Impey}, C.~D., \& {Rix}, H.-W. 2010, \aj, 139,
  2097, \dodoi{10.1088/0004-6256/139/6/2097}

\bibitem[{{Roberts} \& {Haynes}(1994)}]{roberts94}
{Roberts}, M.~S., \& {Haynes}, M.~P. 1994, \araa, 32, 115,
  \dodoi{10.1146/annurev.aa.32.090194.000555}

\bibitem[{{Shen} {et~al.}(2003){Shen}, {Mo}, {White}, {Blanton}, {Kauffmann},
  {Voges}, {Brinkmann}, \& {Csabai}}]{shen03}
{Shen}, S., {Mo}, H.~J., {White}, S. D.~M., {et~al.} 2003, \mnras, 343, 978,
  \dodoi{10.1046/j.1365-8711.2003.06740.x}

\bibitem[{{Shuntov} {et~al.}(2025){Shuntov}, {Akins}, {Paquereau}, {Casey},
  {Ilbert}, {Arango-Toro}, {McCracken}, {Franco}, {Harish}, {Kartaltepe},
  {Koekemoer}, {Yang}, {Huertas-Company}, {Berman}, {McCleary}, {Toft},
  {Gavazzi}, {Achenbach}, {Bertin}, {Brinch}, {Champagne}, {Chartab}, {Drakos},
  {Egami}, {Endsley}, {Faisst}, {Fan}, {Flayhart}, {Hartley}, {Hatamnia},
  {Gozaliasl}, {Gentile}, {Jermann}, {Jin}, {Kakiichi}, {Khostovan},
  {K{\"u}mmel}, {Laigle}, {Laishram}, {Lambrides}, {Liu}, {Lyu}, {Magdis},
  {Mobasher}, {Moutard}, {Renzini}, {Robertson}, {Schefer}, {Scognamiglio},
  {Scoville}, {Sattari}, {Sanders}, {Taamoli}, {Trakhtenbrot}, {Valentino},
  {Wang}, {Weaver}, \& {Yang}}]{shuntov25}
{Shuntov}, M., {Akins}, H.~B., {Paquereau}, L., {et~al.} 2025, arXiv e-prints,
  arXiv:2506.03243, \dodoi{10.48550/arXiv.2506.03243}

\bibitem[{{Skelton} {et~al.}(2014){Skelton}, {Whitaker}, {Momcheva}, {Brammer},
  {van Dokkum}, {Labb{\'e}}, {Franx}, {van der Wel}, {Bezanson}, {Da Cunha},
  {Fumagalli}, {F{\"o}rster Schreiber}, {Kriek}, {Leja}, {Lundgren}, {Magee},
  {Marchesini}, {Maseda}, {Nelson}, {Oesch}, {Pacifici}, {Patel}, {Price},
  {Rix}, {Tal}, {Wake}, \& {Wuyts}}]{skelton14}
{Skelton}, R.~E., {Whitaker}, K.~E., {Momcheva}, I.~G., {et~al.} 2014, \apjs,
  214, 24, \dodoi{10.1088/0067-0049/214/2/24}

\bibitem[{{Smit} {et~al.}(2014){Smit}, {Bouwens}, {Labb{\'e}}, {Zheng},
  {Bradley}, {Donahue}, {Lemze}, {Moustakas}, {Umetsu}, {Zitrin}, {Coe},
  {Postman}, {Gonzalez}, {Bartelmann}, {Ben{\'\i}tez}, {Broadhurst}, {Ford},
  {Grillo}, {Infante}, {Jimenez-Teja}, {Jouvel}, {Kelson}, {Lahav}, {Maoz},
  {Medezinski}, {Melchior}, {Meneghetti}, {Merten}, {Molino}, {Moustakas},
  {Nonino}, {Rosati}, \& {Seitz}}]{smit14}
{Smit}, R., {Bouwens}, R.~J., {Labb{\'e}}, I., {et~al.} 2014, \apj, 784, 58,
  \dodoi{10.1088/0004-637X/784/1/58}

\bibitem[{{Strateva} {et~al.}(2001){Strateva}, {Ivezi{\'c}}, {Knapp},
  {Narayanan}, {Strauss}, {Gunn}, {Lupton}, {Schlegel}, {Bahcall}, {Brinkmann},
  {Brunner}, {Budav{\'a}ri}, {Csabai}, {Castander}, {Doi}, {Fukugita},
  {Gy{\H{o}}ry}, {Hamabe}, {Hennessy}, {Ichikawa}, {Kunszt}, {Lamb}, {McKay},
  {Okamura}, {Racusin}, {Sekiguchi}, {Schneider}, {Shimasaku}, \&
  {York}}]{strateva01}
{Strateva}, I., {Ivezi{\'c}}, {\v{Z}}., {Knapp}, G.~R., {et~al.} 2001, \aj,
  122, 1861, \dodoi{10.1086/323301}

\bibitem[{{Suess} {et~al.}(2022){Suess}, {Bezanson}, {Nelson}, {Setton},
  {Price}, {van Dokkum}, {Brammer}, {Labb{\'e}}, {Leja}, {Miller}, {Robertson},
  {Wel}, {Weaver}, \& {Whitaker}}]{suess22}
{Suess}, K.~A., {Bezanson}, R., {Nelson}, E.~J., {et~al.} 2022, \apjl, 937,
  L33, \dodoi{10.3847/2041-8213/ac8e06}

\bibitem[{{Sun} {et~al.}(2024){Sun}, {Ho}, {Zhuang}, {Ma}, {Chen}, \&
  {Li}}]{sun24}
{Sun}, W., {Ho}, L.~C., {Zhuang}, M.-Y., {et~al.} 2024, \apj, 960, 104,
  \dodoi{10.3847/1538-4357/acf1f6}

\bibitem[{{Szomoru} {et~al.}(2011){Szomoru}, {Franx}, {Bouwens}, {van Dokkum},
  {Labb{\'e}}, {Illingworth}, \& {Trenti}}]{szomoru11}
{Szomoru}, D., {Franx}, M., {Bouwens}, R.~J., {et~al.} 2011, \apjl, 735, L22,
  \dodoi{10.1088/2041-8205/735/1/L22}

\bibitem[{{Trujillo} {et~al.}(2006){Trujillo}, {F{\"o}rster Schreiber},
  {Rudnick}, {Barden}, {Franx}, {Rix}, {Caldwell}, {McIntosh}, {Toft},
  {H{\"a}ussler}, {Zirm}, {van Dokkum}, {Labb{\'e}}, {Moorwood},
  {R{\"o}ttgering}, {van der Wel}, {van der Werf}, \& {van
  Starkenburg}}]{trujillo06}
{Trujillo}, I., {F{\"o}rster Schreiber}, N.~M., {Rudnick}, G., {et~al.} 2006,
  \apj, 650, 18, \dodoi{10.1086/506464}

\bibitem[{{van der Wel} {et~al.}(2008){van der Wel}, {Holden}, {Zirm}, {Franx},
  {Rettura}, {Illingworth}, \& {Ford}}]{vanderWel08}
{van der Wel}, A., {Holden}, B.~P., {Zirm}, A.~W., {et~al.} 2008, \apj, 688,
  48, \dodoi{10.1086/592267}

\bibitem[{{van der Wel} {et~al.}(2012){van der Wel}, {Bell}, {H{\"a}ussler},
  {McGrath}, {Chang}, {Guo}, {McIntosh}, {Rix}, {Barden}, {Cheung}, {Faber},
  {Ferguson}, {Galametz}, {Grogin}, {Hartley}, {Kartaltepe}, {Kocevski},
  {Koekemoer}, {Lotz}, {Mozena}, {Peth}, \& {Peng}}]{vanderWel12}
{van der Wel}, A., {Bell}, E.~F., {H{\"a}ussler}, B., {et~al.} 2012, \apjs,
  203, 24, \dodoi{10.1088/0067-0049/203/2/24}

\bibitem[{{van der Wel} {et~al.}(2014){van der Wel}, {Franx}, {van Dokkum},
  {Skelton}, {Momcheva}, {Whitaker}, {Brammer}, {Bell}, {Rix}, {Wuyts},
  {Ferguson}, {Holden}, {Barro}, {Koekemoer}, {Chang}, {McGrath},
  {H{\"a}ussler}, {Dekel}, {Behroozi}, {Fumagalli}, {Leja}, {Lundgren},
  {Maseda}, {Nelson}, {Wake}, {Patel}, {Labb{\'e}}, {Faber}, {Grogin}, \&
  {Kocevski}}]{vanderWel14}
{van der Wel}, A., {Franx}, M., {van Dokkum}, P.~G., {et~al.} 2014, \apj, 788,
  28, \dodoi{10.1088/0004-637X/788/1/28}

\bibitem[{{van der Wel} {et~al.}(2024){van der Wel}, {Martorano},
  {H{\"a}u{\ss}ler}, {Nedkova}, {Miller}, {Brammer}, {van de Ven}, {Leja},
  {Bezanson}, {Muzzin}, {Marchesini}, {de Graaff}, {Nelson}, {Kriek}, {Bell},
  \& {Franx}}]{vanderWel24}
{van der Wel}, A., {Martorano}, M., {H{\"a}u{\ss}ler}, B., {et~al.} 2024, \apj,
  960, 53, \dodoi{10.3847/1538-4357/ad02ee}

\bibitem[{{van Dokkum} {et~al.}(2011){van Dokkum}, {Brammer}, {Fumagalli},
  {Nelson}, {Franx}, {Rix}, {Kriek}, {Skelton}, {Patel}, {Schmidt}, {Bezanson},
  {Bian}, {da Cunha}, {Erb}, {Fan}, {F{\"o}rster Schreiber}, {Illingworth},
  {Labb{\'e}}, {Lundgren}, {Magee}, {Marchesini}, {McCarthy}, {Muzzin},
  {Quadri}, {Steidel}, {Tal}, {Wake}, {Whitaker}, \& {Williams}}]{vanDokkum11}
{van Dokkum}, P.~G., {Brammer}, G., {Fumagalli}, M., {et~al.} 2011, \apjl, 743,
  L15, \dodoi{10.1088/2041-8205/743/1/L15}

\bibitem[{{Vega-Ferrero} {et~al.}(2024){Vega-Ferrero}, {Huertas-Company},
  {Costantin}, {P{\'e}rez-Gonz{\'a}lez}, {Sarmiento}, {Kartaltepe},
  {Pillepich}, {Bagley}, {Finkelstein}, {McGrath}, {Knapen}, {Arrabal Haro},
  {Bell}, {Buitrago}, {Calabr{\`o}}, {Dekel}, {Dickinson}, {Dom{\'\i}nguez
  S{\'a}nchez}, {Elbaz}, {Ferguson}, {Giavalisco}, {Holwerda}, {Kocesvski},
  {Koekemoer}, {Pandya}, {Papovich}, {Pirzkal}, {Primack}, \&
  {Yung}}]{vega-ferrero24}
{Vega-Ferrero}, J., {Huertas-Company}, M., {Costantin}, L., {et~al.} 2024,
  \apj, 961, 51, \dodoi{10.3847/1538-4357/ad05bb}

\bibitem[{{Ward} {et~al.}(2024){Ward}, {de la Vega}, {Mobasher}, {McGrath},
  {Iyer}, {Calabr{\`o}}, {Costantin}, {Dickinson}, {Holwerda},
  {Huertas-Company}, {Hirschmann}, {Lucas}, {Pandya}, {Wilkins}, {Yung},
  {Arrabal Haro}, {Bagley}, {Finkelstein}, {Kartaltepe}, {Koekemoer},
  {Papovich}, \& {Pirzkal}}]{ward24}
{Ward}, E., {de la Vega}, A., {Mobasher}, B., {et~al.} 2024, \apj, 962, 176,
  \dodoi{10.3847/1538-4357/ad20ed}

\bibitem[{{Williams} {et~al.}(2009){Williams}, {Quadri}, {Franx}, {van Dokkum},
  \& {Labb{\'e}}}]{williams09}
{Williams}, R.~J., {Quadri}, R.~F., {Franx}, M., {van Dokkum}, P., \&
  {Labb{\'e}}, I. 2009, \apj, 691, 1879, \dodoi{10.1088/0004-637X/691/2/1879}

\bibitem[{{Williams} {et~al.}(2010){Williams}, {Quadri}, {Franx}, {van Dokkum},
  {Toft}, {Kriek}, \& {Labb{\'e}}}]{williams10}
{Williams}, R.~J., {Quadri}, R.~F., {Franx}, M., {et~al.} 2010, \apj, 713, 738,
  \dodoi{10.1088/0004-637X/713/2/738}

\bibitem[{{Wuyts} {et~al.}(2012){Wuyts}, {F{\"o}rster Schreiber}, {Genzel},
  {Guo}, {Barro}, {Bell}, {Dekel}, {Faber}, {Ferguson}, {Giavalisco}, {Grogin},
  {Hathi}, {Huang}, {Kocevski}, {Koekemoer}, {Koo}, {Lotz}, {Lutz}, {McGrath},
  {Newman}, {Rosario}, {Saintonge}, {Tacconi}, {Weiner}, \& {van der
  Wel}}]{wuyts12}
{Wuyts}, S., {F{\"o}rster Schreiber}, N.~M., {Genzel}, R., {et~al.} 2012, \apj,
  753, 114, \dodoi{10.1088/0004-637X/753/2/114}

\bibitem[{{Yang} {et~al.}(2025){Yang}, {Kartaltepe}, {Franco}, {Ding},
  {Achenbach}, {Arango-Toro}, {Casey}, {Drakos}, {Faisst}, {Gillman},
  {Gozaliasl}, {Huertas-Company}, {Jin}, {Liu}, {Magdis}, {Massey},
  {Silverman}, {Tanaka}, {Yu}, {Akins}, {Allen}, {Ilbert}, {Koekemoer},
  {McCracken}, {Paquereau}, {Rhodes}, {Robertson}, {Shuntov}, \&
  {Toft}}]{yang25}
{Yang}, L., {Kartaltepe}, J.~S., {Franco}, M., {et~al.} 2025, \apjs, 281, 68,
  \dodoi{10.3847/1538-4365/ae0e1b}

\end{thebibliography}
\bibliographystyle{aasjournal}

\end{document}